\documentclass[showpacs,preprintnumbers,amsmath,amssymb,floatfix]{revtex4}
\usepackage[german,english]{babel}
\usepackage{graphicx}
\usepackage{amsmath}
\usepackage{amsfonts}
\usepackage{amssymb}
\usepackage{epsfig}
\usepackage[hang,nooneline]{subfigure}
\newcommand{\insertplot}[5]{\begin{figure}
 \hfill\hbox to 0.05in{\vbox to #5in{\vfill
 \inputplot{#1}{#4}{#5}}\hfill}
 \hfill\vspace{-.1in}
 \caption{#2}\label{#3}
 \end{figure}}
\newcommand{\inputplot}[3]{
 \special{ps: plotfile #1}
}
 
\newcounter{fig}

\newcommand{\exa}{e^a}
\newcommand{\exddd}{e^{-(D-3)a}}

\begin{document}

\title{\bf Spontaneous Symmetry Breaking in 
Wormholes Spacetimes with Matter}

\author{{\bf Christian Hoffmann $^{1,2}$}}
\email[{\it Email:}]{christian.hoffmann@uni-oldenburg.de}
\author{{\bf Theodora Ioannidou $^3$}}
\email[{\it Email:}]{ti3@auth.gr}
\author{{\bf Sarah Kahlen $^1$}}
\email[{\it Email:}]{sarah.kahlen@uni-oldenburg.de}
\author{{\bf Burkhard Kleihaus $^1$}}
\email[{\it Email:}]{b.kleihaus@uni-oldenburg.de}
\author{{\bf Jutta Kunz $^1$}}
\email[{\it Email:}]{jutta.kunz@uni-oldenburg.de}
\affiliation{
$^1$
Institut f\"ur Physik, Universit\"at Oldenburg, Postfach 2503,
D-26111 Oldenburg, Germany\\
$^2$
Department of Mathematics and Statistics, University of Massachusetts, Amherst,
Massachusetts, 01003-4525, USA\\
$^3$ Department of Mathematics, Physics and Computational Sciences, Faculty of Engineering,\\
Aristotle University of Thessaloniki
Thessaloniki, 54124,  Greece
}

\date{\today}

\begin{abstract}
When bosonic matter in the form of a complex scalar field
is added to Ellis wormholes, the phenomenon of spontaneous symmetry breaking
is observed. Symmetric solutions
possess full reflection symmetry with respect to
the radial coordinate of the two asymptotically flat spacetime regions
connected by the wormhole,
whereas asymmetric solutions do not possess this symmetry.
Depending on the size of the throat,
at bifurcation points pairs of asymmetric solutions
arise from or merge with the symmetric solutions.
These asymmetric solutions are energetically favoured.
When the backreaction of the boson field is taken into account, 
this phenomenon is retained. 
Moreover, in a certain region of the solution space
both symmetric and asymmetric solutions exhibit a transition
from single throat to double throat configurations.
\end{abstract}

\maketitle

\section{Introduction}

Spontaneous symmetry breaking is a ubiquitous phenomenon in physics.
Its wide range of applications includes,
for instance, the Higgs mechanism in particle physics, 
which allows to give mass to the particles of the Standard Model, 
or the phase transitions of ferromagnets in solid state physics.
Here we consider this phenomenon in gravity.

In particular, we consider an Ellis wormhole 
in General Relativity 
\cite{Ellis:1973yv,Ellis:1979bh,Bronnikov:1973fh,Morris:1988cz,Morris:1988tu,ArmendarizPicon:2002km,Sushkov:2005kj,Lobo:2005us}. 
Such a wormhole
connects two asymptotically flat spacetimes by a throat. 
In order to allow for the non-trivial topology in General Relativity
a phantom field is included, i.e., a real scalar field with a
reversed sign in front of its kinetic term in the action.

When symmetric boundary conditions are specified,
the Ellis wormhole is reflection symmetric with respect to its throat.
In suitable coordinates, the wormhole metric in $D=4$ spacetime dimensions reads
\begin{equation}
ds^2 = - dt^2 + d \eta^2 + \left(\eta^2 + \eta_0^2 \right) d \Omega^2 \ .
\label{ellis_sym}
\end{equation}
Then the wormhole throat is located at $\eta=0$,
and the throat parameter $\eta_0$ characterizes the size of the throat.
Of course, by choosing asymmetric boundary conditions also
asymmetric wormhole solutions can be created,
where the reflection symmetry $\eta \to - \eta$ no longer holds.
However, here we are not interested in such an enforced symmetric breaking.
Instead we would like to demonstrate that asymmetric wormholes can appear
also for symmetric boundary conditions, when matter is included.
Thus the symmetry breaking happens spontaneously, and energy considerations
will tell us, that the asymmetric solutions are energetically preferred.

Wormholes immersed in matter have been studied before with
various types of matter. Examples include nuclear matter
\cite{Dzhunushaliev:2011xx,Dzhunushaliev:2012ke,Charalampidis:2013ixa} 
and bosonic matter \cite{Dzhunushaliev:2014bya}.
However, most solutions studied so far were symmetric,
and the asymmetric solutions were asymmetric by construction,
because of an asymmetric choice of boundary conditions.
Thus - to our knowledge - for wormhole solutions immersed in matter
spontaneous symmetry breaking has not been observed before.

Recently, the static Ellis wormhole solution has
been generalized to spacetimes with higher dimensions 
\cite{Torii:2013xba,Dzhunushaliev:2013jja}.
Moreover, rotating generalizations of the Ellis wormhole 
in four and five dimensions
have been found \cite{Dzhunushaliev:2013jja,Kleihaus:2014dla,Chew:2016epf}.
We will not address the question of rotation here.
Instead we will consider non-rotating wormhole solutions in $D$ dimensions
immersed in bosonic matter, with a focus on five dimensions.

The bosonic matter consists of a complex scalar field with a
self-interaction such that it allows for localized solutions
already in a flat spacetime background, which correspond to
non-topological solitons \cite{Friedberg:1976me,Friedberg:1986tq} or $Q$-balls
\cite{Coleman:1985ki}.
When gravity is coupled to this bosonic matter boson stars arise
\cite{Friedberg:1986tq,Lee:1991ax}, which can get very close
to the black hole limit.
In \cite{Dzhunushaliev:2014bya} in addition a phantom field was included,
to obtain solutions with a non-trivial topology, i.e.,
$Q$-balls and boson stars harbouring wormholes at their core.
However, all these solutions were reflection symmetric.

Here we study the emergence of the asymmetric solutions.
Let us denote with ${\cal M}_+$ the part of the manifold
with positive radial coordinate $\eta$, and with ${\cal M}_-$
the part with negative $\eta$. Since we are dealing with
configurations with two asymptotically flat regions,
we obtain two values for the mass, $M_\pm$,
as measured asymptotically in ${\cal M}_\pm$.
These values can be read off the asymptotic behaviour
of the metric. For symmetric solutions both masses agree. However,
for asymmetric solutions the two masses can differ widely.
We note, that the asymmetric solutions always come in pairs, 
since they are transformed into each other
by the reflection transformation $\eta \to -\eta$.
Consequently, their two masses are simply interchanged.

To obtain the particle number
in the case of $Q$-balls and boson stars,
volume integrals are performed.
For the solutions with
non-trivial topology one can proceed
analogously, when the solutions are symmetric. For the 
asymmetric solutions, however, the calculation of the
particle number via such integrals can become ambiguous,
since the inner boundary is not provided by symmetry.
Therefore we here propose an unambiguous procedure
to obtain the particle number. 
This is mandatory,
since we need the particle number in order to demonstrate that
the asymmetric solutions are energetically favourable.

The presence of the matter in the wormhole spacetime has further notable
consequences. In particular, the backreaction of the matter
on the geometry can cause a drastic change of the
geometry, giving rise to a transition 
from configurations with a single throat to 
configurations with a double throat and an equator in between.
Such a transition is known to occur for symmetric wormholes
with bosonic matter \cite{Dzhunushaliev:2014bya}
and for other types of matter as well
\cite{Hauser:2013jea,Dzhunushaliev:2014mza}.

This paper is organized as follows:
In section II we present the theoretical setting
for obtaining both symmetric and asymmetric
configurations of wormholes immersed in bosonic matter
in $D$ dimensions.
The numerical results are shown in section III, 
starting with the probe limit,
where the spontaneous symmetry breaking is already observed.
We then discuss in detail the five-dimensional families
of symmetric and asymmetric solutions in the presence of gravity.
Finally, we include a brief discussion
of the asymmetric configurations in four dimensions,
and show that the spontaneous symmetry breaking is present as well.
We conclude in section V.
In appendix A we briefly address our method of extracting
the mass and the particle number for asymmetric solutions.

\section{Theoretical setting}

\subsection{Action}

We consider General Relativity with a minimally coupled complex scalar field $\Phi$
and phantom field $\Psi$ in $D$ spacetime dimensions.
Besides the Einstein-Hilbert action
with curvature scalar $\cal R$, coupling constant $\kappa$ 
and metric determinant $g$, the action 
\begin{equation}
S=\int \left[ \frac{1}{2 \kappa}{\cal R} + 
{\cal L}_{\rm ph} +{\cal  L}_{\rm bs} \right] \sqrt{-g}\  d^Dx  
 \label{action}
\end{equation}
then contains the respective matter contributions,
the Lagrangian ${\cal L}_{\rm ph}$ of the phantom field $\Psi$,
and the Lagrangian ${\cal L}_{\rm bs}$ of the complex scalar field $\Phi$.
The kinetic term for the phantom field  carries the reverse sign
\begin{equation}
 {\cal L}_{\rm ph} = \frac{1}{2}\partial_\mu \Psi\partial^\mu \Psi \ ,
\label{lpsi}
\end{equation}
as compared to the kinetic term of the complex scalar field
\begin{equation}
{\cal L}_{\rm bs} = 
-\frac{1}{2} g^{\mu\nu}\left( \partial_\mu\Phi^* \partial_\nu\Phi
                            + \partial_\nu\Phi^* \partial_\mu\Phi 
 \right) - U( \left| \Phi \right|) \ .
\label{lphi}
\end{equation} 
The asterisk denotes complex conjugation,
and $U$ denotes the potential with the mass term and the self-interaction
\begin{equation}
U(|\Phi|) =  \lambda |\Phi|^2 \left( |\Phi|^4 -c|\Phi|^2 +b \right)
\ . \label{U} 
\end{equation}
The global minimum of the potential resides at $\Phi =0$, where $U(0)=0$, 
while a local minimum is found at some finite value of $|\Phi|$.
The mass of the bosons $m_b=\sqrt{\lambda b}$ is specified by the
quadratic term.
The potential is chosen such that it allows for non-topological soliton solutions
\cite{Friedberg:1976me,Friedberg:1986tq} or $Q$-balls
\cite{Coleman:1985ki}.

Variation of the action with respect to the metric
leads to the Einstein equations
\begin{equation}
G_{\mu\nu}= {\cal R}_{\mu\nu}-\frac{1}{2}g_{\mu\nu}{\cal R} =  \kappa T_{\mu\nu}
\label{ee} 
\end{equation}
with stress-energy tensor
\begin{equation}
T_{\mu\nu} = g_{\mu\nu}{{\cal L}}_M
-2 \frac{\partial {{\cal L}}_M}{\partial g^{\mu\nu}} \ ,
\label{tmunu} 
\end{equation}
where we denoted by ${\cal L}_{\rm M} = 
{\cal L}_{\rm ph}+{\cal L}_{\rm bs} $ the sum of the scalar field Lagrangians.

\subsection{Ans\"atze}

For the line element 
of the spherically symmetric solutions
with a non-trivial topology we choose
%
%
\begin{equation}
ds^2 = -e^{(D-3)a} dt^2 
    +p e^{-a}\left[d\eta^2 + \left(\eta^2 + \eta_0^2\right)\Omega^2_{D-2}\right] \ .
\label{lineel}
\end{equation}
Here $d\Omega^2_{D-2}$ 
denotes the metric on the unit $(D-2)$-sphere, while $a$ and $p$ are functions of 
the radial coordinate $\eta$,
which takes positive and negative 
values, i.e. $-\infty< \eta < \infty$. 
The two limits $\eta\to \pm\infty$
correspond to two distinct asymptotically flat regions,
associated with ${\cal M}_+$ and ${\cal M}_-$, respectively.
Note that in Eq.~(\ref{lineel}) we have introduced the parameter $\eta_0$, 
which we will refer to as the throat parameter.

As for spherically symmetric $Q$-balls and boson stars, 
we parametrize the complex scalar field $\Phi$ via
\begin{equation}
\Phi 
  =  \phi (\eta) ~ e^{ i\omega_s t } \ ,   \label{ansatzp}
\end{equation}
where $\phi (\eta)$ is a real function,
and $\omega_s$ denotes the boson frequency.
The phantom field $\Psi$ depends only on the radial coordinate,
\begin{equation}
\Psi 
=  \psi (\eta) \ .
\label{ansatzph}
\end{equation}

\subsection{Einstein and Matter Field Equations}

Substituting the above Ans\"atze into the Einstein equations 
$G_\mu^\nu=\kappa  T_\mu^\nu$ leads to the following field equations
\begin{eqnarray}
-\frac{D-2}{2p}\exa (a''-p''/p)
+\frac{(D-2)(D-3)}{8p}\exa (a'^2-2 a' p'/p)  & & \nonumber\\
+\frac{(D-2)(D-7)}{8p^3}\exa p'^2
-\frac{(D-2)^2}{2ph}\exa \eta ( a'-p'/p) & & \nonumber\\
-\frac{(D-2)(D-5)}{2 p h^2}\exa \eta_0^2
 & = &
-\kappa \left(U(\phi) +\frac{\exa}{2p}(2\phi'^2-\psi'^2) +\omega_s^2 e^{-(D-3)a}\phi^2\right)
\label{eeq_00}
\end{eqnarray}
\begin{eqnarray}
-\frac{(D-2)(D-3)}{8p}\exa a'^2
+\frac{(D-2)(D-3)}{8p^3}\exa p'^2 & & \nonumber\\
+\frac{(D-2)(D-3)}{2p^2 h}\exa \eta p'
-\frac{(D-2)(D-3)}{2 p h^2}\exa \eta_0^2
 & = &
-\kappa \left(U(\phi) -\frac{\exa}{2p}(2\phi'^2-\psi'^2) -\omega_s^2 e^{-(D-3)a}\phi^2\right)
\label{eeq_rr}
\end{eqnarray}
\begin{eqnarray}
 \frac{D-3}{2p^2}\exa p''
+\frac{(D-2)(D-3)}{8p}\exa a'^2& & \nonumber\\
+\frac{(D-3)(D-8)}{8p^3}\exa p'^2
+\frac{(D-3)^2}{2p^2 h}\exa \eta p'& & \nonumber\\
-\frac{(D-3)(D-6)}{2 p h^2}\exa \eta_0^2
& = &
-\kappa \left(U(\phi) +\frac{\exa}{2p}(2\phi'^2-\psi'^2) -\omega_s^2 e^{-(D-3)a}\phi^2\right)
\label{eeq_tt}
\end{eqnarray}
which derive from the $tt$, $\eta\eta$ and $\theta\theta$ components, respectively.
For convenience we use the abbreviation $h=\eta^2+\eta_0^2$.

Variation of the action with respect to the
complex scalar field and to the phantom field 
leads to the equations
\begin{eqnarray}
\phi'' +\left(\frac{D-3}{2}\frac{p'}{p}+(D-2)\frac{\eta}{h}\right) \phi' 
& = & 
\frac{1}{2} p e^{-a} \frac{dU}{d\phi}-\omega_s^2 p  e^{-(D-2)a}\phi \ ,
\label{eqSk}\\
\left(\left(ph\right)^{\frac{D-2}{2}}\frac{1}{\sqrt{p}}\psi'\right)' & = & 0 \ ,
\label{eqPh}
\end{eqnarray}
where integration of the last equation leads to
\begin{equation}
\psi' = \sqrt{p}\left(ph\right)^{-\frac{D-2}{2}}{\cal D}\ .
\label{phip}
\end{equation}
Here the constant $\cal D$ denotes the scalar charge of the phantom field.
By inserting Eq.~(\ref{phip}) into Eq.~(\ref{eeq_rr})
the scalar charge $\cal D$ 
can be expressed via
\begin{eqnarray}
{\cal D}^2 
  = \left(ph\right)^{D-2}\left[
 \frac{(D-2)(D-3)}{4p} 
 \left(a'^2 - p'^2/p^2-4\frac{\eta}{h} p'/p +4\frac{\eta_0^2}{h^2}\right)
 -2\kappa e^{-a}\left(U(\phi) -\frac{\exa}{p}\phi'^2 -\omega_s^2 \phi^2 \exddd\right)
 \right] \ .
\label{eqD2}
\end{eqnarray}
%

%
We now eliminate the $\phi'^2$ term 
and the $\psi'^2$ term in the Einstein equations
by adding Eq.~(\ref{eeq_rr}) to Eq.~(\ref{eeq_00}) 
and to Eq.~(\ref{eeq_tt}).
This provides us with 
the final set of Einstein equations
\begin{eqnarray}
a''
+\frac{D-2}{h}a'
-\frac{D-3}{2p} \eta a' p'
 & = & -4\kappa \frac{p e^{-a}}{(D-2)(D-3)}\left(U(\phi)-\omega_s^2(D-2) \phi^2 \exddd\right)
\label{eqa}\\
p''+
\frac{(2 D-5)}{h}\eta p'
+\frac{D-5}{2p}  p'^2
-2\frac{(D-4)\eta_0^2}{h^2}p
 & = & -4\kappa \frac{p^2 e^{-a}}{D-3}\left(U(\phi)-\omega_s^2 \phi^2 \exddd\right)
\label{eqp}
\end{eqnarray}
to be solved 
together with the equation for the bosonic matter, Eq.~(\ref{eqSk}). 
Clearly, the system of equations allows for reflection symmetric
solutions ${\cal S}$, i.e., solutions whose functions are either symmetric
or antisymmetric under $\eta \to - \eta$.

\subsection{Boundary Conditions}

We need to solve the above set of three coupled ordinary differential equations
of second order. Thus we have to impose six boundary conditions.
Here we would like to impose symmetric boundary conditions, which are the same
for $\eta \to \infty$ and $\eta \to -\infty$. Therefore any asymmetry in the
solutions is not enforced via boundary conditions but arises spontaneously.

The boundary conditions for the boson field function $\phi(\eta)$
are then given by
\begin{equation}
\phi(\eta\to \pm \infty) \to 0 \ .
\label{bcasym}
\end{equation}
These conditions ensure, that the configurations possess finite energy.

For the metric we require asymptotic flatness in both asymptotic regions.
By imposing on the metric function $a$ the conditions
\begin{equation}
a(\eta\to \pm \infty) \to 0 \  
\label{bcbsym}
\end{equation}
we also set the time scale.
For the metric function $p$ asymptotic flatness in both asymptotic regions
implies
\begin{equation}
p(\eta\to \pm \infty) \to 1 \  .
\label{bccsym}
\end{equation}
%

\subsection{Single and multiple throats}

Here we discuss the throat properties in terms of the 
circumferential function $R(\eta)= \sqrt{p h}e^{-a/2}$, which represents
the radius of a circle in the equatorial plane with constant coordinate
$\eta$.

Let us first consider symmetric solutions ${\cal S}$.
In symmetric solutions, the metric functions are symmetric under
$\eta \to - \eta$.
Consequently, $\eta=0$ plays a special role.
The metric functions $a$ and $p$ possess a vanishing derivative at $\eta=0$,
$a'(0)=p'(0)=0$, which implies that they assume extremal values.
In particular, from the circumferential radius $R$ we can conclude,
that if $R(0)$ is a minimum, a throat is located at $\eta=0$,
since $\eta=0$ then corresponds to a minimal surface.

If $R(0)$ is a maximum, it represents only a local maximum,
since $R(\eta) \to |\eta|$  asymptotically. Such a local maximum
corresponds to a maximal surface and thus an equator.
Clearly, between the equator and asymptotic infinity (at least) one
throat should be localized in each of the regions ${\cal M}_+$ and ${\cal M}_-$.
In that case double (or multiple) throat configurations are present.

Transitions between single and double throat configurations
in four spacetime dimensions were observed for symmetric solutions in
\cite{Dzhunushaliev:2014bya}.
These transitions occur, when the second and first derivative of $R$ vanish,
\begin{equation}
R''(0) = 0 \ \ \Longleftrightarrow \ \ \ 
\left[ (D-2)(D-3)\exa - 2 p \eta_0^2 \kappa U(\phi) \right]_{\eta=0} =0 \ . 
\end{equation}
%
%
We note that for symmetric solutions the surface gravity vanishes at 
the throat for configurations with a single
throat, since the surface gravity is given by 
$\frac{D-3}{2}\left[ e^{(D-2)a/2} a'/\sqrt{p}\right]_{\eta_{\rm th}}$,
and $a'(0)=0$ 
\cite{Dzhunushaliev:2014bya}.

In contrast to symmetric solutions, where a single throat
must be located at $\eta=0$, a throat of asymmetric solutions 
can be located anywhere. Moreover, the transition from 
single to double throat solutions can happen, when the
circumferential radius develops a turning point 
at some value of the radial coordinate $\eta_{0\, \rm cr}$, where
$R'(\eta_{0\, \rm cr})=R''(\eta_{0\, \rm cr})=0$.
Thus for asymmetric solutions the single throat does not degenerate
to form an equator and a double throat, but an equator together with
a second throat appear spontaneously somewhere else
in the manifold.

\subsection{Energy conditions}

When the null energy condition (NEC) is violated,
also the weak and the strong energy condition are violated.
It is therefore sufficient to address only the NEC,
which requires
\begin{equation}
\Xi = T_{\mu\nu} k^\mu k^\nu \ge 0 \ 
\end{equation}
for all (future-pointing) null vector fields $k^\mu$.

This condition can be expressed via the Einstein tensor
by making use of the Einstein equations.
For spherically symmetric solutions we then obtain 
the new conditions
\begin{equation}
-G_t^t+G_\eta^\eta \geq  0 \ , \ \ \ {\rm and } 
\ \ \ -G_t^t+G_\theta^\theta\geq  0 \ ,
\label{Nulleng}
\end{equation}
both of which must be obeyed everywhere in order to respect
the NEC.
For the present set of solution the NEC is always violated.

\subsection{Mass and scalar charge}

In the presence of gravity, the mass of a
stationary asymptotically flat solution in $D$ dimensions
can be obtained from the Komar integral \cite{Wald:1984} 
\begin{equation}
M = - \frac{1}{16 \pi G_D} \frac{D-2}{D-3} \int_{S_{\infty}^{D-2}} \alpha \ ,
 \label{komarM}
\end{equation}
with $\alpha_{\mu_1 \dots \mu_{D-2}} \equiv \epsilon_{\mu_1 \dots \mu_{D-2}
  \rho \sigma} \nabla^\rho \xi^\sigma$ and $\xi \equiv \partial_t$.
By choosing the surface $S$ at spatial infinity in ${\cal M}_+$ (${\cal M}_-$)
we obtain the mass $M_+$ ($M_-$).
Both values of the mass are encoded in the metric function $g_{tt}$
and can easily be extracted. We obtain $M_+$ from
\begin{equation}
g_{tt} \underset{\eta \to \infty}{\longrightarrow} - 1 + \frac{\mu}{\eta^{D-3}} \ , \ \ \
M_+= \frac{(D-2) \Omega_{D-2}}{16 \pi G} \mu
\ , \label{komarM2}
\end{equation}
where $\Omega_{D-2}$ is the area of the unit $D-2$-sphere,
and $M_-$ analogously. When we use Stokes' theorem to convert the integral
into a volume integral, we obtain a boundary term at some $\eta_b$, e.g.,
\begin{equation}
{M_+} = 
 \frac{1}{{4\pi}} \int_{\Sigma}
 R_{\mu\nu}n^\mu\xi^\nu dV  
- \frac{1}{16 \pi G_D} \frac{D-2}{D-3} \int_{S_{\eta_b}^{D-2}} \alpha \
=  \frac{1}{{4\pi}} \int_{\Sigma}
 R_{\mu\nu}n^\mu\xi^\nu dV   + M_{\eta_b}
\ . \label{komarM1}
\end{equation}
Here $\Sigma$ denotes an asymptotically flat spacelike hypersurface,
$n^\mu$ is normal to $\Sigma$ with $n_\mu n^\mu = -1$,
and $dV$ is the natural volume element on $\Sigma$
\cite{Wald:1984}.
Obviously, the volume integral only agrees
with the mass $M_+$ when the surface term $M_{\eta_b}$ vanishes.
Since the surface term is proportional to the product of the surface gravity and the
surface area, this is the case when the surface gravity vanishes.

In the symmetric case, the surface gravity always vanishes at $\eta=0$,
i.e., at the throat (or at the equator).
Then the volume integral may be evaluated to give the mass.
However, in the asymmetric case, the surface gravity is finite at the
throat (and at the equator) \cite{Dzhunushaliev:2014bya}.
In that case the surface term would have to be included to obtain
the proper value of the mass from the volume integral.
A simple evaluation of the volume integral from say $\eta=0$ to 
$\eta=\infty$ alone would not yield the correct value for the mass 
in the asymmetric case. These considerations are important
for evaluating the mass of the asymmetric configurations in the
probe limit. Therefore we present an appropriate procedure for extracting the
mass in the probe limit in Appendix A.

\subsection{Charge or particle number}

The Lagrange density is invariant under the global phase transformation
\begin{equation}
\displaystyle
\Phi \rightarrow \Phi e^{i\chi} \ .
\end{equation}
This leads to the conserved current
\begin{eqnarray}
j^{\mu} & = &  - i \left( \Phi^* \partial^{\mu} \Phi 
 - \Phi \partial^{\mu}\Phi ^* \right) \ , \ \ \
j^{\mu} _{\ ; \, \mu}  =  0 \ .
\label{current}
\end{eqnarray}
In globally regular topologically trivial spacetimes
the associated conserved charge $Q$ is then obtained 
by integrating the time-component
of the current over the entire space
\begin{eqnarray}
Q &= &- \int j^t \left| g \right|^{1/2} d\eta d\Omega_{D-2}
\nonumber \\
&=& 2 \Omega_{D-2} \omega \int_0^{\infty} |g| ^{1/2}   \frac{\phi^2}{A^2} \,d\eta \, \ . 
\label{Qc}
\end{eqnarray}
The global charge $Q$ then corresponds to the particle number
of the complex boson field.

In wormhole spacetimes, we have to reconsider the above
definition of the charge or particle number.
For symmetric configurations with a single throat it is clear that we should integrate
from $\eta=0$ to $\pm \infty$, to obtain the particle numbers $Q_\pm$ 
for the regions ${\cal M}_\pm$. This should remain true, when
the throat at $\eta=0$ turns into an equator \cite{Dzhunushaliev:2014bya}.
However, in the asymmetric case it is no longer obvious, where the inner
integration boundary $\eta_b$ should reside.
If we were to retain $\eta_b$ at the throat, for instance, 
an ambiguity would arise
at the very least when the second throat and the equator emerge.

To solve this question, we reconsider the case of the mass. 
In the presence of gravity,
the mass is obtained unambiguously and in a simple way, 
when we use surface integrals at plus and minus infinity.
If we were to use surface integrals for the charge or particle number as well,
we would also have a clean definition of the charges $Q_\pm$ 
to be associated with the masses $M_\pm$.
To achieve this feat we employ the following trick.
We minimally couple the complex scalar field to a 
fictitious U(1) gauge field, which is not allowed to backreact
on the configuration.
Thus the configuration is not changed, while we can employ 
the Gauss law to read off the charge $Q_\pm$ of the configurations
in the asymptotically flat regions
\begin{equation}
Q_\pm = \int_{S_{\pm \infty} }   {^*F} .
\label{chargeNew}
\end{equation}
This definition then yields the same charge as the above volume integral in the symmetric case,
while it leads to meaningful and unambiguous values in the asymmetric case,
which can be compared with the respective masses 
to extract the binding energies of these configurations.

\section{Wormholes immersed in bosonic matter}

When solving the field equations subject to the given set 
of symmetric boundary conditions, we were in for a surprise,
since the numerical procedure led to asymmetric solutions
in addition to the expected symmetric solutions.
Our further studies then revealed the phenomenon of
spontaneous symmetry breaking in wormhole spacetimes
with bosonic matter. The demonstration of this 
phenomenon is the main focus of this section.

While we have presented the formalism in the last section
for $D$ spacetime dimensions, 
we have performed the numerical calculations 
mainly in five dimensions,
although we have also performed a study in four dimensions to
convince ourselves that the analogous 
features are present.

In this section we first address the probe limit,
where the boson field equation is solved in the
background of the Ellis wormhole.
Already here the spontaneous symmetry breaking
is observed.
Depending on the throat size,
the asymmetric solutions (${\cal A}_\pm$) can bifurcate
from the symmetric solutions (${\cal S}$)
at critical values of the boson frequency.
The asymmetric solutions are always energetically favoured.

Subsequently, we couple to gravity, and thus take
the backreaction of the boson field into account.
The phenomenon of spontaneous symmetry breaking
in then retained.
However, the structure of the families of solutions
becomes much richer. In particular, both
symmetric and asymmetric solutions exhibit transitions
to double throat wormhole configurations.

Finally, we address this phenomenon in four dimensions,
were contact to astrophysics can in principle be made.
Here the symmetric solutions were studied in detail
before, but the asymmetric ones had not been seen
because the calculations had been restricted to ${\cal M}_+$
\cite{Dzhunushaliev:2014bya}.

In the following we focus on the fundamental solutions. 
We refer to solutions as fundamental
when they do not possess a radially excited boson field,
i.e., when their boson field function does not have nodes.
However, we have observed that the
radially excited solutions exhibit the analogous pattern
of symmetric and asymmetric solutions.

For the numerical calculations we have introduced
the compactified  radial coordinate
\begin{equation}
x= {\rm atan}\left(\frac{\eta}{r_0}\right) \ ,
\end{equation}
where $r_0$ is some constant, for which we chose $r_0=3$.
We have then employed a collocation method for boundary-value ordinary
differential equations, equipped with an adaptive mesh selection procedure
\cite{COLSYS}.
We have used typical mesh sizes with $10^3-10^4$ points,
reaching a relative accuracy of $10^{-10}$ for the functions.
Estimates of the relative errors for the mass and the angular momentum
have been of order $10^{-6}$.
We have employed the condition ${\cal D}=\mbox{const}$, Eq.~(\ref{eqD2}), 
to monitor the quality of the numerical solutions.
The variation of ${\cal D}$ has been typically less than $10^{-9}$.

For the self-interaction potential $U(\phi)$ we have chosen
the parameters $\lambda=1$, $c=2$ and $b=1.1$,
since this allowed us to compare with previous
calculations of bosonic configurations without a wormhole
\cite{Hartmann:2010pm}.
With these potential parameters fixed, the further parameters were
the parameter $\kappa$, which includes the
gravitational coupling strength,
the boson frequency $\omega_s$,
and the throat parameter $\eta_0$.
In most of the results shown, we have fixed the throat parameter $\eta_0=3$,
leaving only $\kappa$ and $\omega_s$ as free parameters.

\subsection{Probe Limit}


We here demonstrate, that the phenomenon of
spontaneous symmetry breaking occurs already
in the probe limit.
We begin by illustrating the symmetric and asymmetric solutions,
and then discuss their dependence on the two parameters,
the boson frequency $\omega_s$ and the
throat size $\eta_0$. We point out that depending on the throat size
bifurcations occur.
Subsequently, we consider the 
masses and particle numbers of the configurations
and show that the asymmetric solutions
are more strongly bound.

\subsubsection{Parameters}

\begin{figure}[t!]
\begin{center}
\vspace{0.5cm}
\mbox{\hspace{-0.5cm}
\subfigure[][]{\hspace{-1.0cm}
\includegraphics[height=.25\textheight, angle =0]{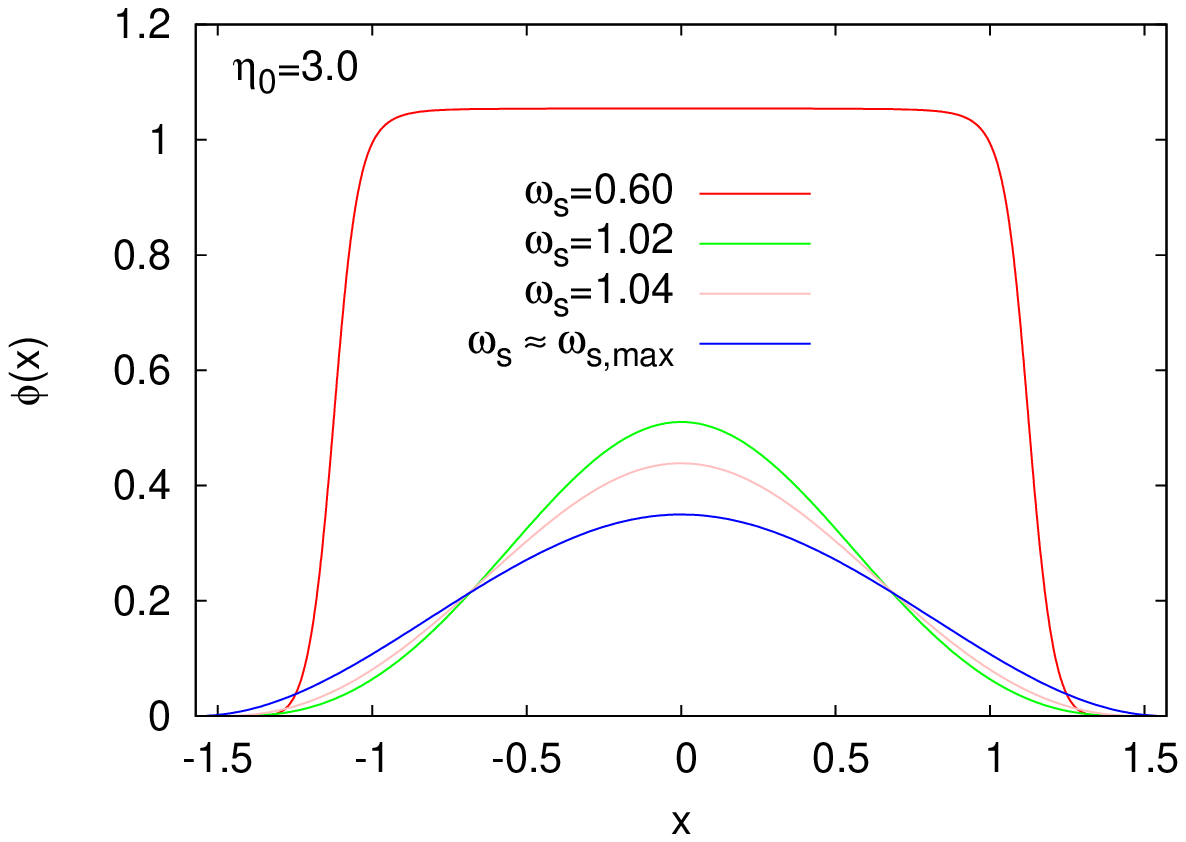}
\label{fig1a}
}
\subfigure[][]{\hspace{-0.5cm}
\includegraphics[height=.25\textheight, angle =0]{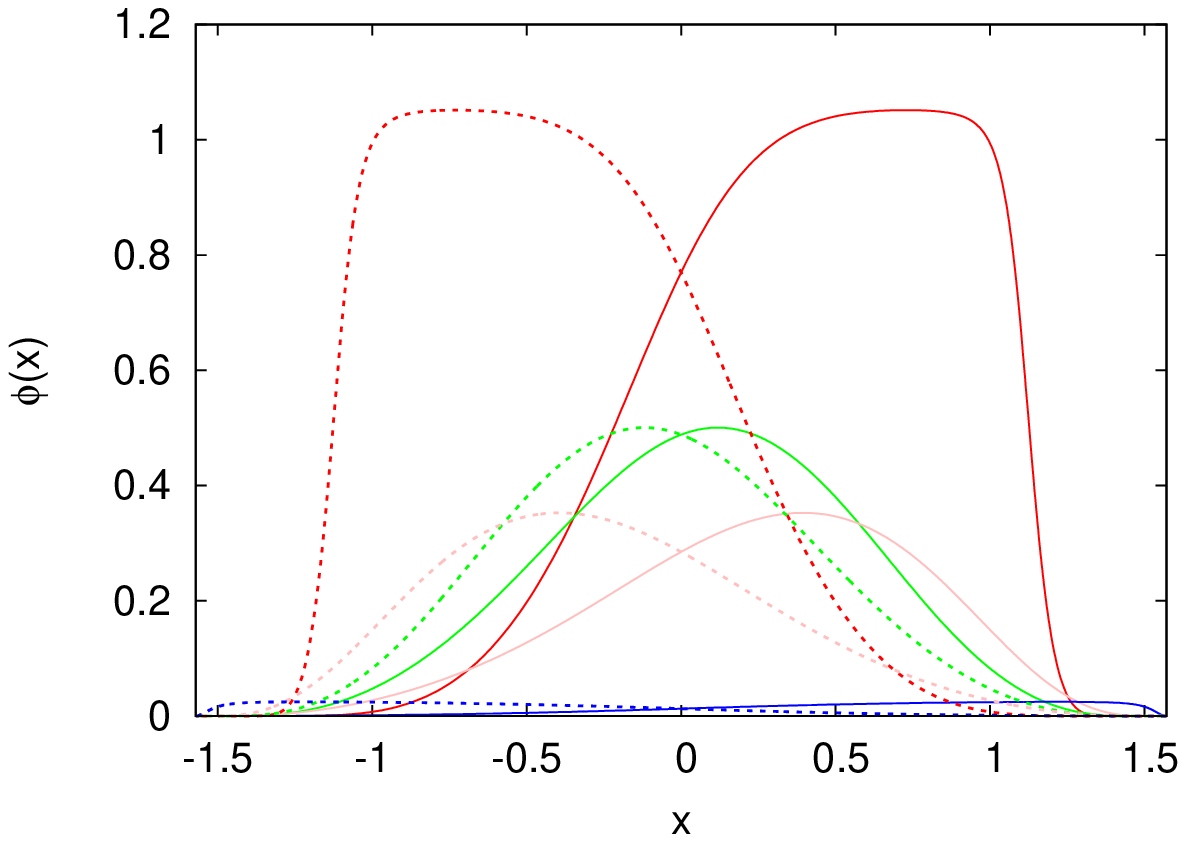}
\label{fig1b}
}
}
\mbox{\hspace{-0.5cm}
\subfigure[][]{\hspace{-1.0cm}
\includegraphics[height=.25\textheight, angle =0]{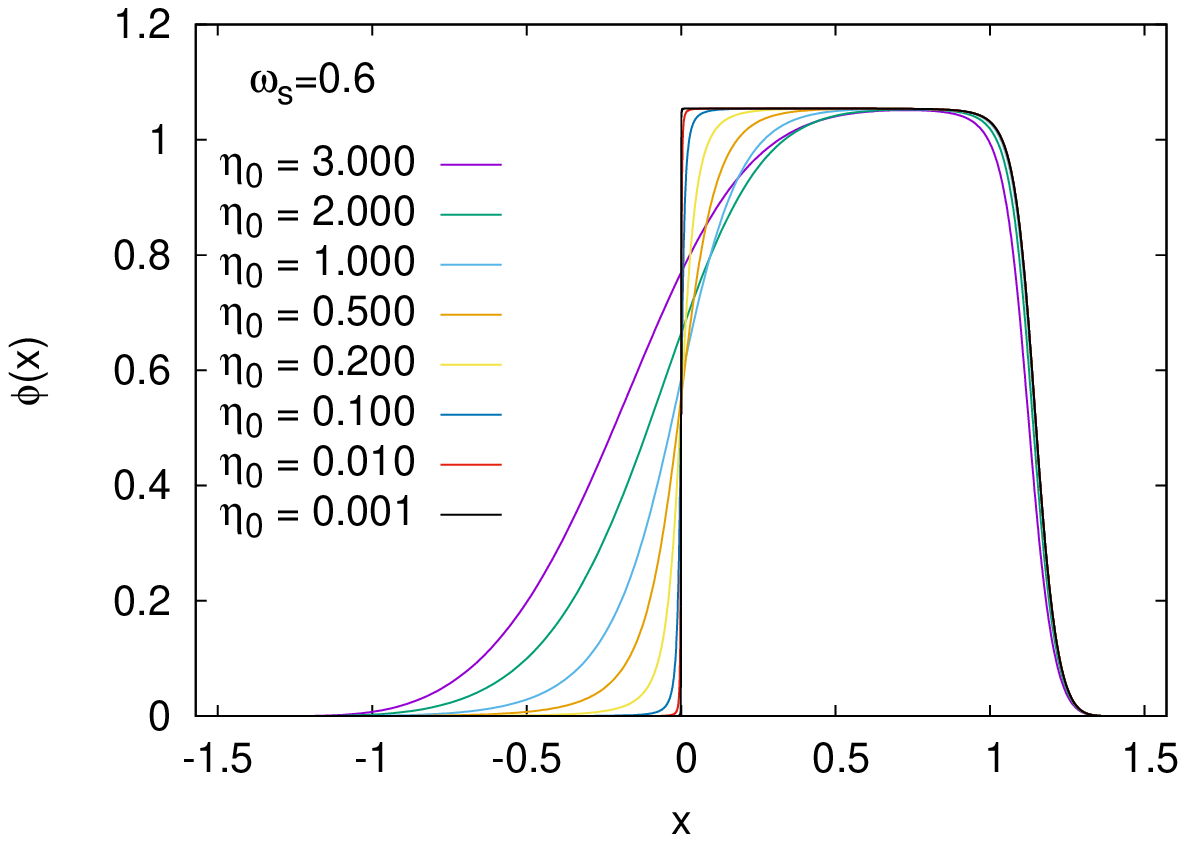}
\label{fig1c}
}
\subfigure[][]{\hspace{-0.5cm}
\includegraphics[height=.25\textheight, angle =0]{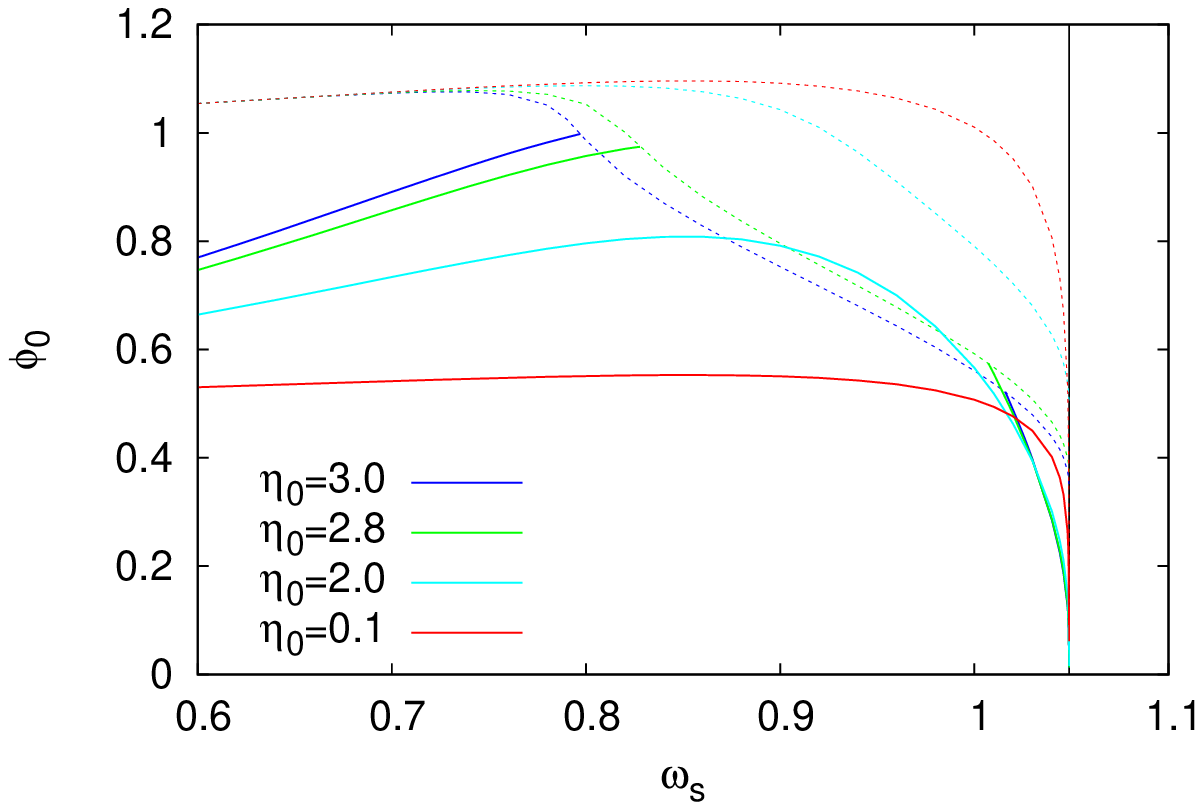}
\label{fig1d}
}
}
\mbox{\hspace{-0.5cm}
\subfigure[][]{\hspace{-1.0cm}
\includegraphics[height=.25\textheight, angle =0]{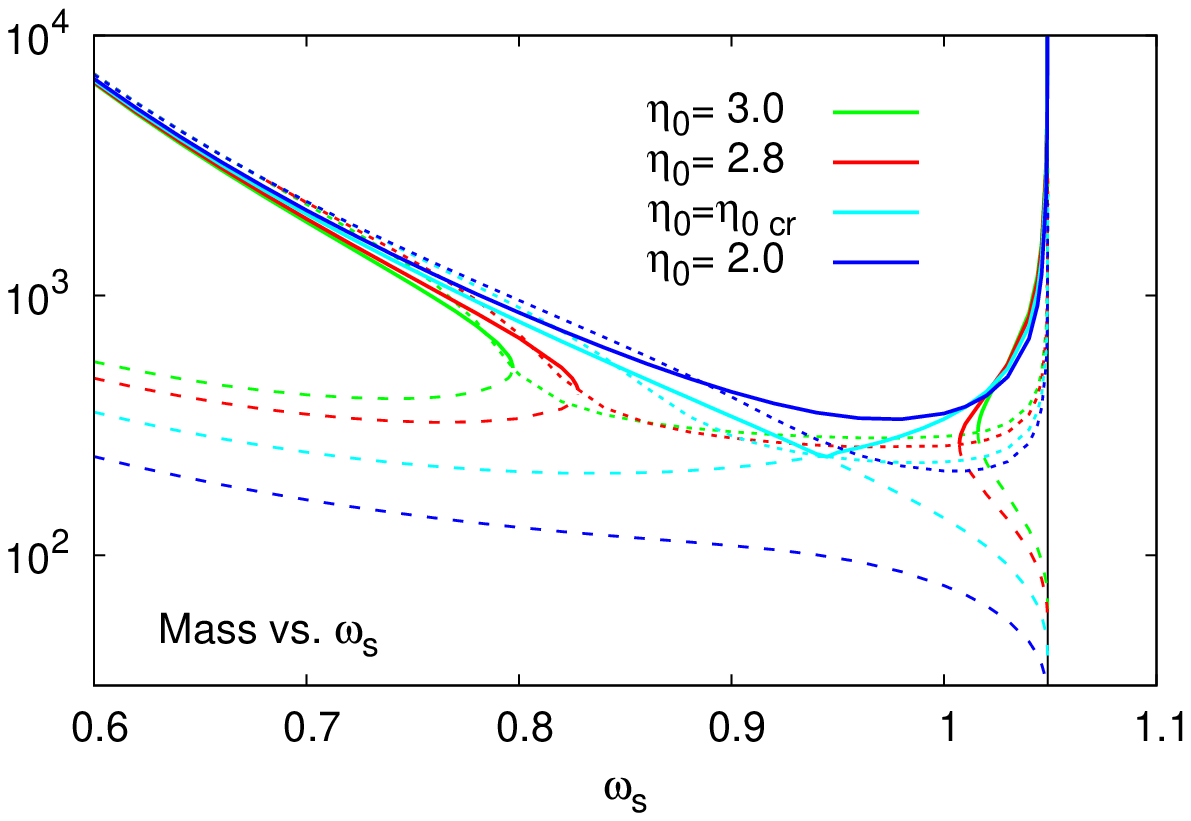}
\label{fig1e}
}
\subfigure[][]{\hspace{-0.5cm}
\includegraphics[height=.25\textheight, angle =0]{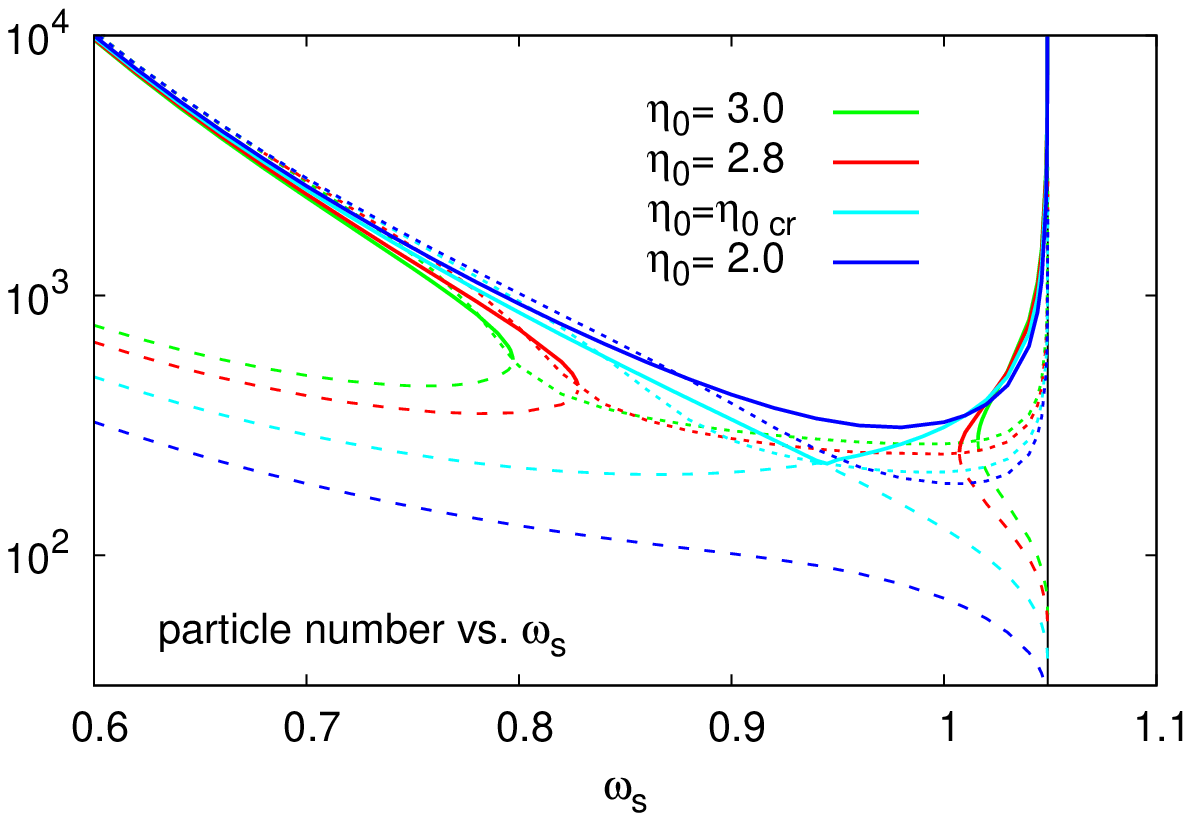}
\label{fig1f}
}
}
\end{center}
\vspace{-0.7cm}
\caption{Probe limit:
(a) The boson field function $\phi$ versus the
compactified coordinate $x={\rm atan}\, (\eta/r_0)$
for a fixed throat parameter $\eta_0=3.0$
and decreasing values of the boson frequency
$\omega_s$.
(b) Same as (a) for asymmetric solutions.
(c) The boson field function $\phi$ versus the
compactified coordinate $x={\rm atan}\, (\eta/r_0)$
for asymmetric solutions at a fixed boson frequency
$\omega_s=0.6$ and decreasing values of the
throat parameter $\eta_0$.
(d) The value $\phi_0$ of the boson field functions $\phi$ 
at the throat $\eta=0$ versus the boson frequency
$\omega_s$ for decreasing values of the
throat parameter $\eta_0$.
(e) The mass $M$ (dotted) of the symmetric solutions 
and the mass $M_+$ (solid) and $M_-$ (dashed)
of the asymmetric solutions versus the boson frequency
$\omega_s$ for decreasing values of the
throat parameter $\eta_0$.
(f) Same as (e) for the particle number.
\label{fig1}
}
\end{figure}

Usually the probe limit is obtained by simply taking the coupling $\kappa$
of gravity and matter to zero. However, since we want to
retain the non-trivial topology,
we must be careful here as to take only the coupling to the complex
boson field to zero, while retaining the coupling to the phantom field.
This can be achieved by an appropriate scaling of the phantom field,
$\Psi \to \Psi/\sqrt{\kappa}$.
The Einstein equations then yield the Ellis wormhole 
\cite{Torii:2013xba,Dzhunushaliev:2013jja},
and only the boson field equation needs to be solved
in the background of the Ellis wormhole.

The solutions depend on
the boson frequency $\omega_s$ and on the throat parameter $\eta_0$.
In the limit $\eta_0 \to 0$, contact must be made with
topologically trivial solutions, since the two asymptotically flat
regions are then separated.
Indeed, for $\eta_0 \to 0$, the so called 
non-topological soliton or $Q$-ball limit is reached
\cite{Friedberg:1976me,Coleman:1985ki}.
However, it can either be reached in both regions 
${\cal M}_+$ and ${\cal M}_-$ at the same time (symmetric case),
or only in a single region (asymmetric case), as we show below.

The domain of existence of
non-rotating $Q$-balls is restricted to a certain frequency range,
$\omega_{\rm min} < \omega_s < \omega_{\rm max}$
\cite{Friedberg:1976me,Coleman:1985ki,Volkov:2002aj}.
The maximal frequency $\omega_{\rm max}$
\begin{equation}
\label{cond1}
\omega^2_{\rm max} =
\frac{1}{2} U''(0) = \lambda \, b = m_b^2 \ ,
\end{equation}
ensures that the solutions possess an exponential fall-off
at spatial infinity.
The minimal frequency $\omega_{\rm min}$ is based on 
an argument to allow for localized solutions
\cite{Friedberg:1976me,Coleman:1985ki,Volkov:2002aj,Hartmann:2010pm}
and given by
\begin{equation}
\label{cond2}
\omega^2_{\rm min} =
\min_{\phi} \left[{U(\phi)}/{\phi^2} \right] \; = \; 
\lambda \left(b- \frac{c^2}{4} \right) \ .
\end{equation}
These limits are retained for the topologically non-trivial solutions,
where the Minkowski background is replaced by an Ellis background
\cite{Dzhunushaliev:2014bya}.

\subsubsection{Bifurcations}

In Fig.~\ref{fig1} we show features of the symmetric  solutions in the probe limit which are analogous to those
known in four dimensions \cite{Dzhunushaliev:2014bya}.
Here the boson function $\phi$ is symmetric with respect to 
the reflection $\eta \to - \eta$, as seen in Fig.~\ref{fig1a},
where the throat parameter $\eta_0$ is fixed while 
the boson frequency $\omega_s$ is varied.
Note, that as the boson frequency $\omega_s$ tends to $\omega_{\rm min}$,
the boson function $\phi$ tends to a constant in an increasingly
large region.

We show the boson function $\phi$ of the asymmetric solutions
for the same parameters in Fig.~\ref{fig1b}.
Here the reflection symmetry of the solutions is broken.
However, since the field equation is reflection symmetric, 
all asymmetric solutions come in pairs. 
To distinguish the two solutions of a pair, let us include an index $\pm$
associated with the two regions ${\cal M}_\pm$.
The index $+$ ($-$) then indicates that more matter is localized 
in the region ${\cal M}_+$ (${\cal M}_-$).
The functions $\phi_-$ are obtained from
the functions $\phi_+$ by reflection ($\eta \to - \eta$)
\begin{equation}
\label{reflection} \phi_-(\eta) = \phi_+(-\eta) \ ,
\end{equation}
and vice versa.

Let us now vary the throat parameter $\eta_0$ while keeping
the frequency $\omega_s$ fixed.
In the symmetric case, when $\eta_0 \to 0$,
a {\sl double}
$Q$-ball solution is approached, where 
a $Q$-ball is localized in each of the regions
${\cal M}_+$ and ${\cal M}_-$, which
simply represent two disjunct
Minkowski spacetimes in the limit.
In contrast, in the asymmetric case for fixed frequency
and $\eta_0 \to 0$,
the solutions approach a {\sl single}
$Q$-ball, that is 
localized in one of the regions, either
${\cal M}_+$ for $\phi_+$ or ${\cal M}_-$ for $\phi_-$, while the 
complementary region becomes completely empty in the limit.
This is demonstrated in Fig.~\ref{fig1c}.

To monitor the different solutions, it is useful to keep track of
the value of the boson function at the throat, $\phi_0$.
This parameter then maps out the domain of existence of
the symmetric and asymmetric solutions
concerning the dependence on the frequency $\omega_s$
and on the throat parameter $\eta_0$.
We note, that both asymmetric solutions possess the same
value of $\eta_0$.
To illustrate this dependence,
we exhibit in Fig.~\ref{fig1d} the value of $\phi_0$ versus the
boson frequency $\omega_s$ for four values of the
throat parameter, $\eta_0=3$, 2.8, 2 and 0.1.
Recall, that for $\eta_0=0$ the non-trivial topology is lost.

The figure shows that a bifurcation phenomenon must take place
at a critical value $\eta_{0\, \rm cr}$ of the throat parameter between 2.8 and 2,
that is associated with a critical value of the 
boson frequency $\omega_{\rm cr}$.
For $\eta_0 \le  \eta_{0\, \rm cr}$ symmetric and asymmetric
solutions exist in the full interval $[\omega_{\rm min}, \omega_{\rm max}]$.
At the critical value $\eta_{0\, \rm cr}$, symmetric and asymmetric solutions
precisely touch at the critical value $\omega_{\rm cr}$ of the boson frequency.
For $\eta_0 > \eta_{0\, \rm cr}$ a frequency gap 
$\omega_{\rm crl} \le \omega_s \le \omega_{\rm cru}$
appears, where only symmetric solutions exist. 
The pairs of asymmetric solutions then bifurcate 
from the symmetric ones at the end points of the gap, i.e., at the upper
critical frequency $\omega_{\rm cru}$
and at the lower critical frequency $\omega_{\rm crl}$.
The asymmetric solutions
then persist for $\omega_{\rm min} < \omega_s <\omega_{\rm crl}$
and $\omega_{\rm cru} < \omega_s < \omega_{\rm max}$.

We note that the value of the boson field at the throat, $\phi_0$, 
tends to a limiting value, $\phi_0(\omega_{\rm min})=1$ in the symmetric case,
when $\omega_s \to \omega_{\rm min}$.
Here the field equation is solved by $\phi(\eta)=1$.
At first glance it may be surprising, that for small $\eta_0$
($\eta_0=0.1$ in the figure) the value
of $\phi_0$ of the asymmetric solutions is much lower than
the corresponding value of the symmetric solutions,
while both approach a $Q$-ball solution in the limit
$\eta_0 \to 0$. The reason for this is, that in the asymmetric case,
in the limit $\eta_0 \to 0$
the $\phi_+$ field has to jump from its maximal value to zero
at $\eta=0$ (and likewise the $\phi_-$ field).
Therefore for sufficiently small values of 
$\eta_0$ the asymmetric field assumes about half its maximal value 
at $\eta=0$.

\subsubsection{Mass and particle number}

We now turn to the global charges of these solutions, beginning with the mass.
In Fig.~\ref{fig1e} we exhibit the mass
versus the boson frequency $\omega_s$
for several values of the throat parameter $\eta_0$,
including the critical value $\eta_{0\, \rm cr}$.
The mass $M$ of the symmetric solutions is shown by dotted curves,
the mass $M_+$ of the solutions with more matter localized
in ${\cal M}_+$ is represented by solid curves, and
the mass $M_-$ of the solutions with more matter localized
in ${\cal M}_-$ by dashed curves.
We recall that these masses
refer to the asymptotic behaviour in ${\cal M_+}$.
Because of the symmetry of the pair of asymmetric solutions,
the values of their masses would be interchanged 
when read off in ${\cal M_-}$.

The figure nicely illustrates the bifurcation phenomenon seen already in 
Fig.~\ref{fig1d}.
Below $\eta_{0\, \rm cr}$ there are three distinct curves
for $M$, $M_+$ and $M_-$ in the full frequency range.
At $\eta_{0\, \rm cr}$ the three curves touch at the critical 
boson frequency $\omega_{\rm cr}$,
while beyond $\eta_{0\, \rm cr}$ there are two 
bifurcation points of the boson frequency, 
$\omega_{\rm crl}$ and $\omega_{\rm cru}$,
where the pairs of asymmetric solutions bifurcate from 
the symmetric ones.
Away from such bifurcation points, the mass $M_-$ is much smaller
than the other masses.

The particle number $Q$ is exhibited in Fig.~\ref{fig1f}.
Employing the same style for the respective $Q$ curves as for the $M$ 
curves, we see, that the dependence of the particle number on the
throat size and on the boson frequency is completely analogous to the
dependence of the mass.

\begin{figure}[t!]
\begin{center}
\vspace{0.5cm}
\mbox{\hspace{-0.5cm}
\subfigure[][]{\hspace{-1.0cm}
\includegraphics[height=.25\textheight, angle =0]{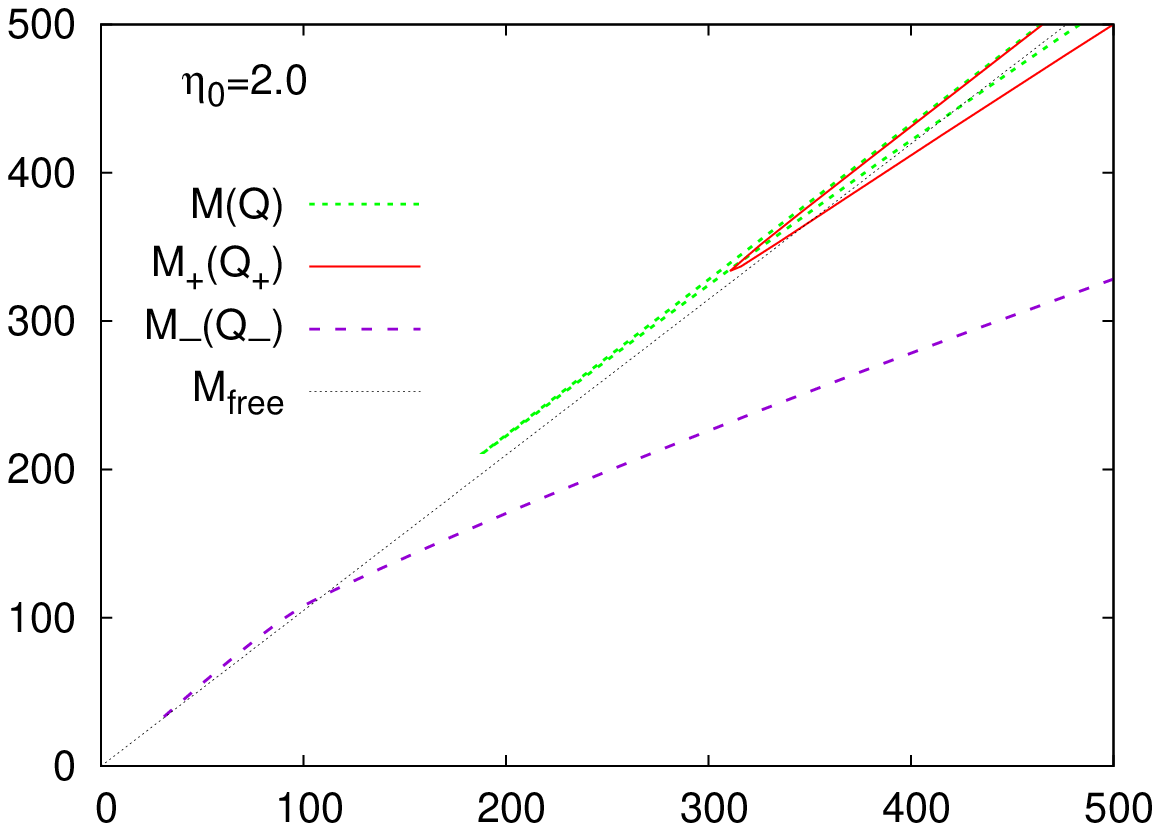}
\label{fig2a}
}
\subfigure[][]{\hspace{-0.5cm}
\includegraphics[height=.25\textheight, angle =0]{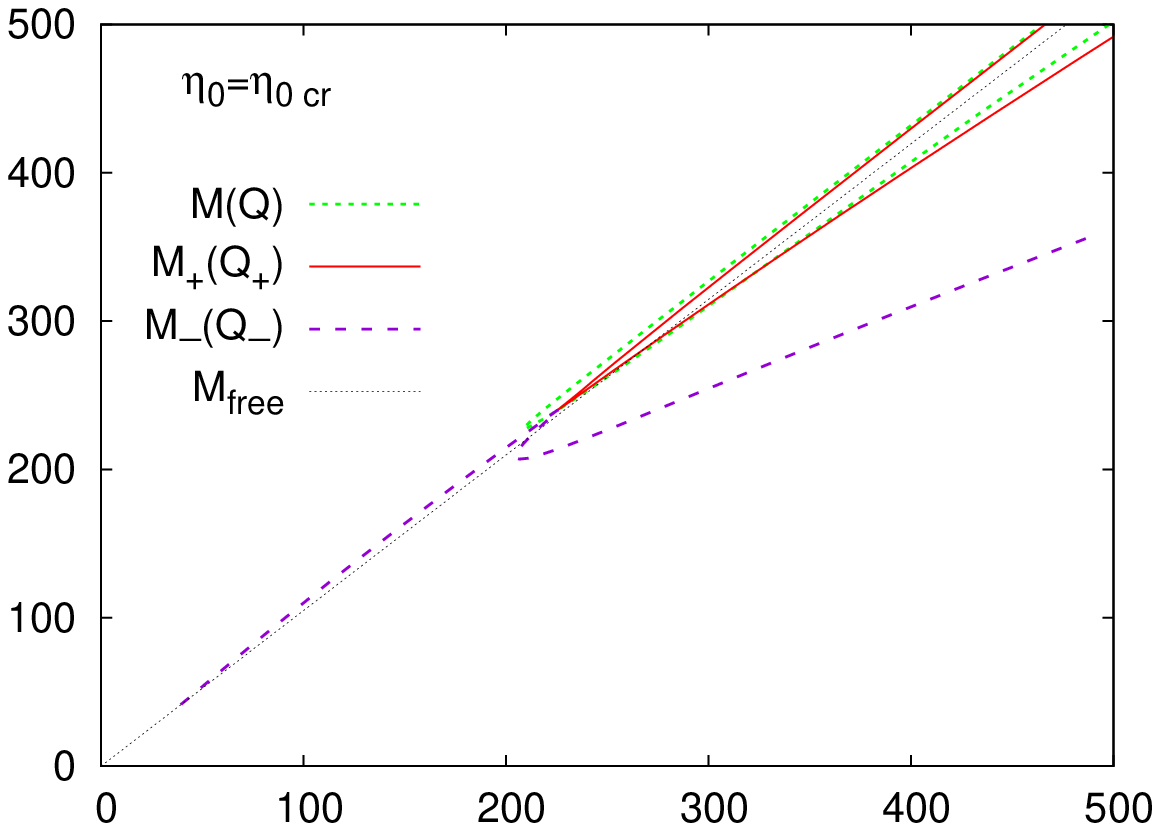}
\label{fig2b}
}
}
\mbox{\hspace{-0.5cm}
\subfigure[][]{\hspace{-1.0cm}
\includegraphics[height=.25\textheight, angle =0]{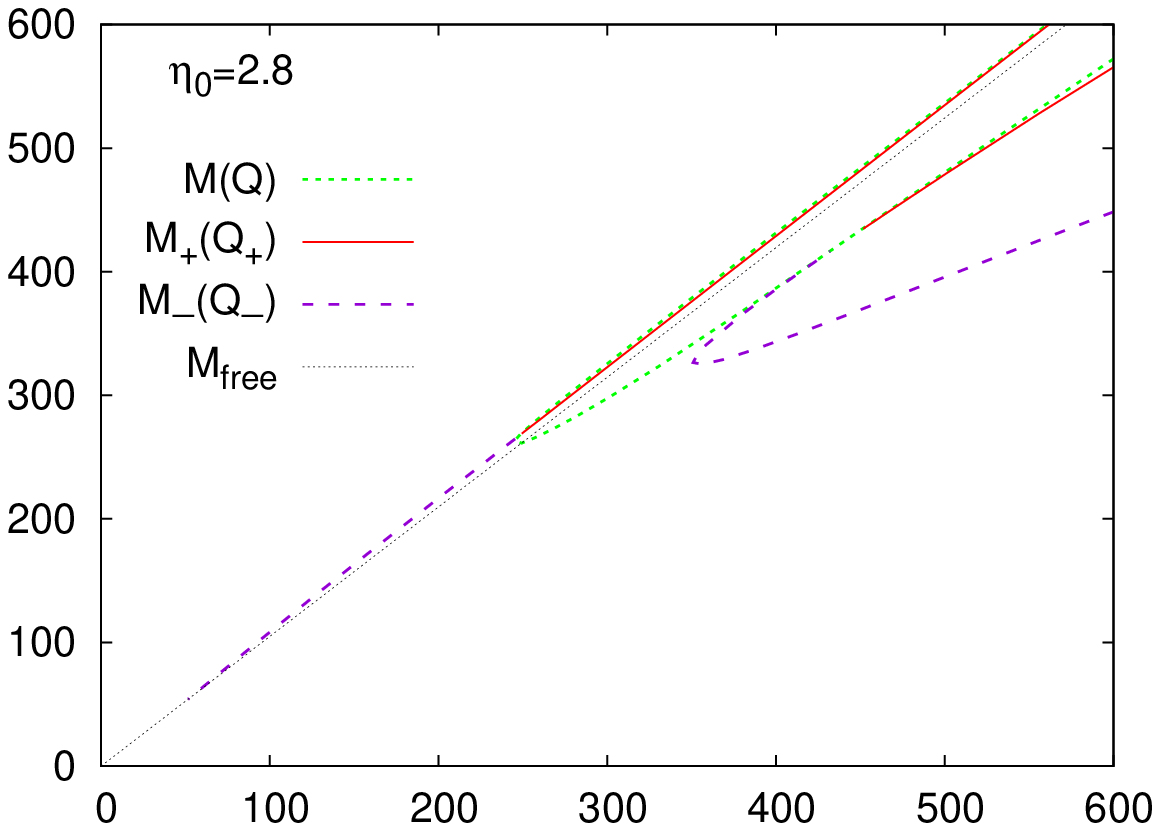}

\label{fig2c}
}
\subfigure[][]{\hspace{-0.5cm}
\includegraphics[height=.25\textheight, angle =0]{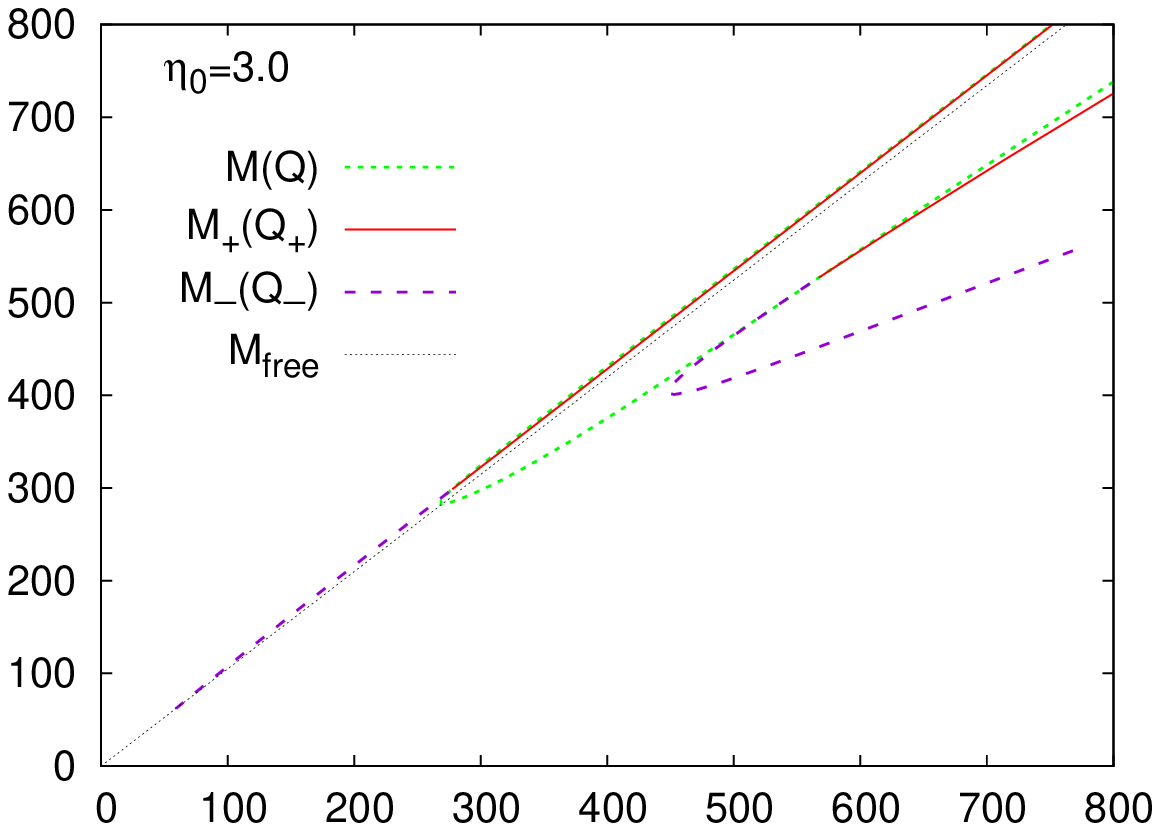}
\label{fig2d}
}
}
\end{center}
\vspace{-0.7cm}
\caption{Probe limit:
(a) The mass $M$ (dotted) of the symmetric solutions
and the mass $M_+$ (solid) and $M_-$ (dashed)
of the asymmetric solutions versus the respective
particle number $Q$, $Q_+$, and $Q_-$
for throat parameter $\eta_0=2$ in ${\cal M}_+$.
(b) Same as (a) for the critical value of $\eta_0$.
(c) Same as (a) for $\eta_0=2.8$.
(d) Same as (a) for $\eta_0=3$.
The mass of $Q$ free bosons is also shown (thin-dotted).
\label{fig2}
}
\end{figure}

Le us now demonstrate that due to spontaneous symmetry breaking 
the asymmetric solutions are energetically favourable.
To this end we address the question, in which of the solutions
the bosons are most strongly bound.
The mass of $Q$ free bosons is given by
\begin{equation}
M_{\bf free} = m_b \, Q \ .
\end{equation}
The binding energy of solutions with $Q$ ($Q_\pm$) 
particles can then be extracted
from the difference of their mass $M$ ($M_\pm$) 
and the mass of $Q$ ($Q_\pm$) free bosons.

When considering $M(Q)$, $M_+(Q_+)$ and $M_-(Q_-)$
one obtains cusp-like structures,
as illustrated in Fig.~\ref{fig2}, where
the masses $M$, $M_+$ and $M_-$ are shown 
versus their respective particle numbers
for the same set of throat parameters as in Fig.~\ref{fig1e}.
Also shown in the figure is the mass of $Q$ free bosons.

The symmetric solutions always form a cusp at some minimal
value of the mass and the particle number.
From this cusp two branches emerge, where the lower branch
soon becomes bound, i.e., $M<M_{\bf free}$, whereas the
upper branch remains unbound.

The branch structure of the asymmetric solutions depends
on the throat parameter and the bifurcation phenomenon.
For $\eta_0 < \eta_{0\, \rm cr}$, the mass $M_+$ of the
asymmetric solutions possesses an analogous cusp structure
to the symmetric one. However beyond $\eta_{0\, \rm cr}$
the respective upper and lower branch are no longer
connected.
Then one clearly notices the bifurcations at $\omega_{\rm crl}$
and $\omega_{\rm cru}$, where in each case two asymmetric branches
bifurcate from a symmetric branch.

In the vicinity of each bifurcation point, both of the emerging asymmetric solutions
possess a lower mass for the same particle number than
the respective symmetric solutions. Thus they are energetically favoured,
and they remain energetically favoured,
also far from their respective bifurcation points.
As expected, the spontaneous symmetry breaking 
leads to energetically more favourable solutions.

\subsection{Gravitating Solutions}

We now consider the backreaction of the boson field on the wormhole.
This means that the full set of coupled nonlinear Einstein-matter equations is solved.
Then in addition to the throat parameter $\eta_0$ and the
boson frequency $\omega_s$ the coupling constant to 
gravity, which is contained in the parameter $\kappa$, 
enters as another continuous parameter.
To reduce the resulting amount of data, we here fix the throat
parameter to $\eta_0=3$, a value above the critical value.
This will allow us to see, how the coupling to gravity
affects the symmetry breaking.
Again we will focus on solutions in five dimensions.

Fixing the values of $\kappa$ and $\eta_0$ (as well as the number
of dimensions), we then
obtain families of gravitating solutions,
formed again for both symmetric (${\cal S}$) and asymmetric (${\cal A}_\pm$)
solutions, which depend on the boson frequency $\omega_s$.
However, unlike the case of the probe limit, the dependence of these
families of solutions on $\omega_s$ can become very involved
in the presence of gravity. Moreover, the wormhole geometry changes
at certain frequencies from single throat to double throat configurations.

\subsubsection{Symmetry Breaking}

\begin{figure}[t!]
\begin{center}
\vspace{0.5cm}
\mbox{\hspace{-0.5cm}
\subfigure[][]{\hspace{-1.0cm}
\includegraphics[height=.25\textheight, angle =0]{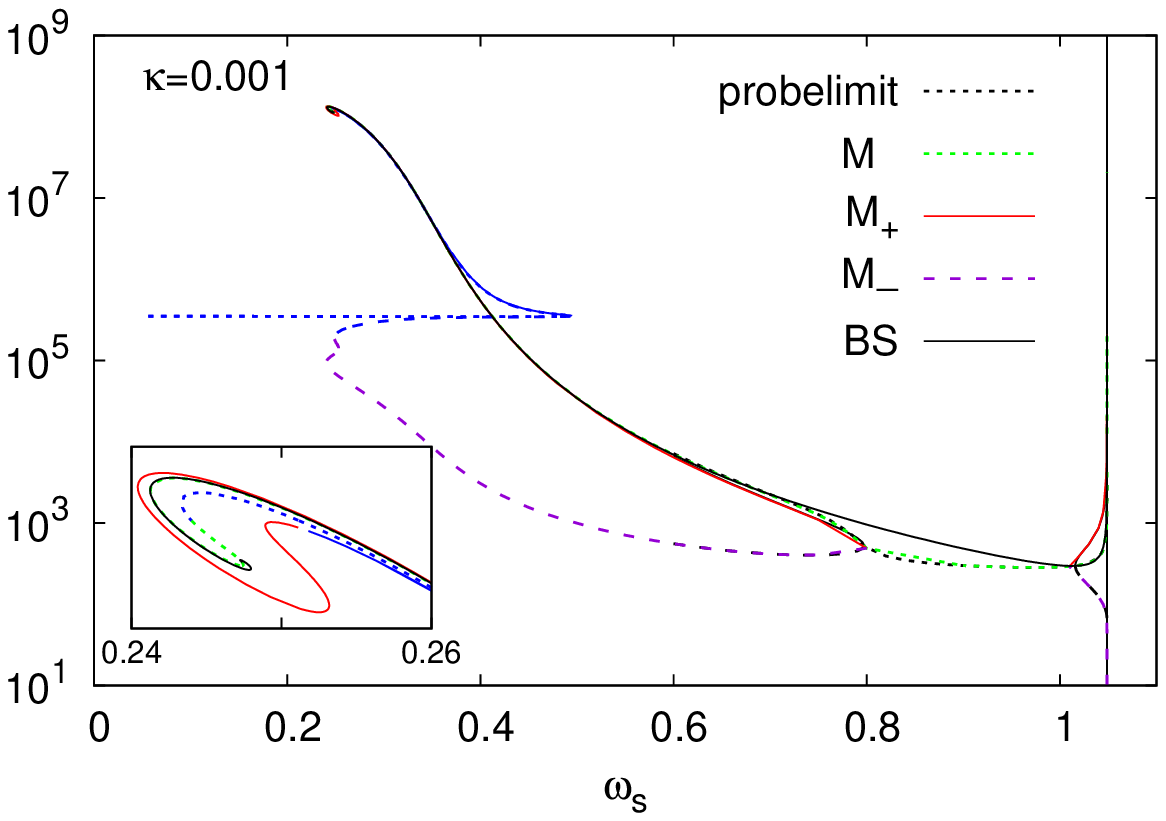}
\label{fig3a}
}
\subfigure[][]{\hspace{-0.5cm}
\includegraphics[height=.25\textheight, angle =0]{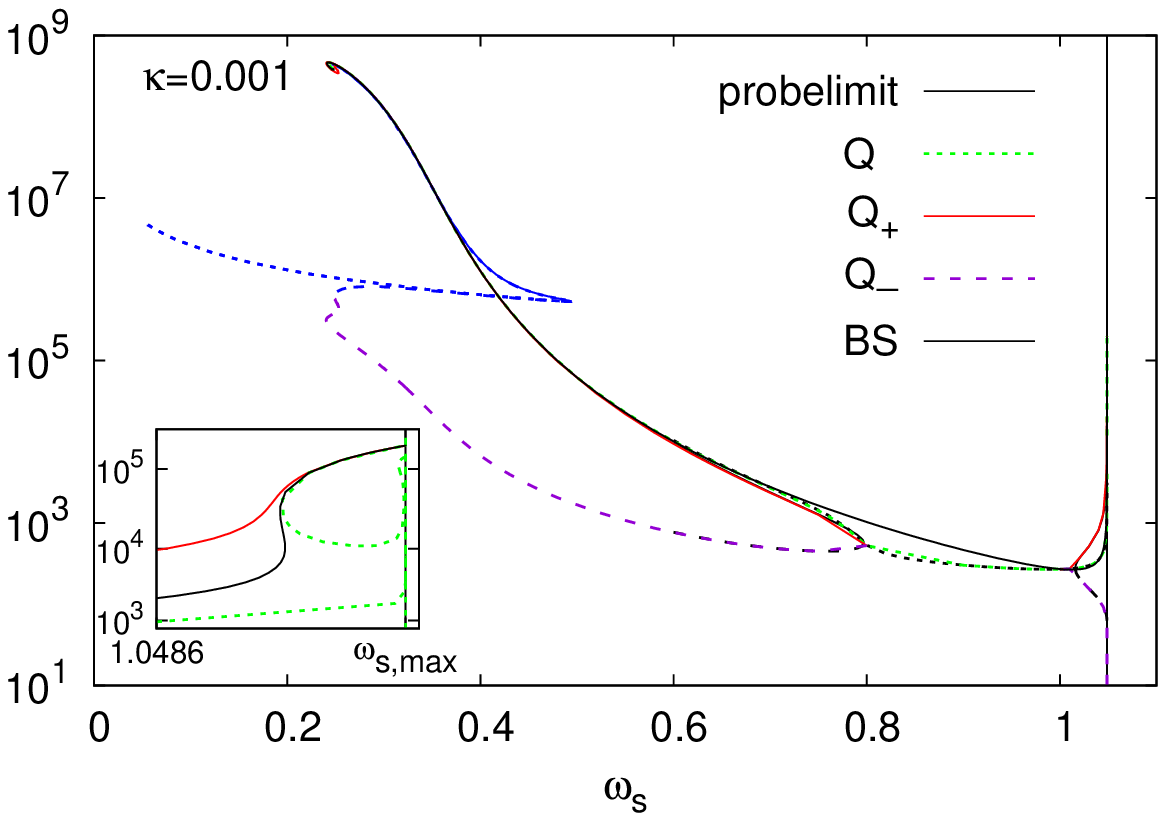}
\label{fig3b}
}
}
\mbox{\hspace{-0.5cm}
\subfigure[][]{\hspace{-1.0cm}
\includegraphics[height=.25\textheight, angle =0]{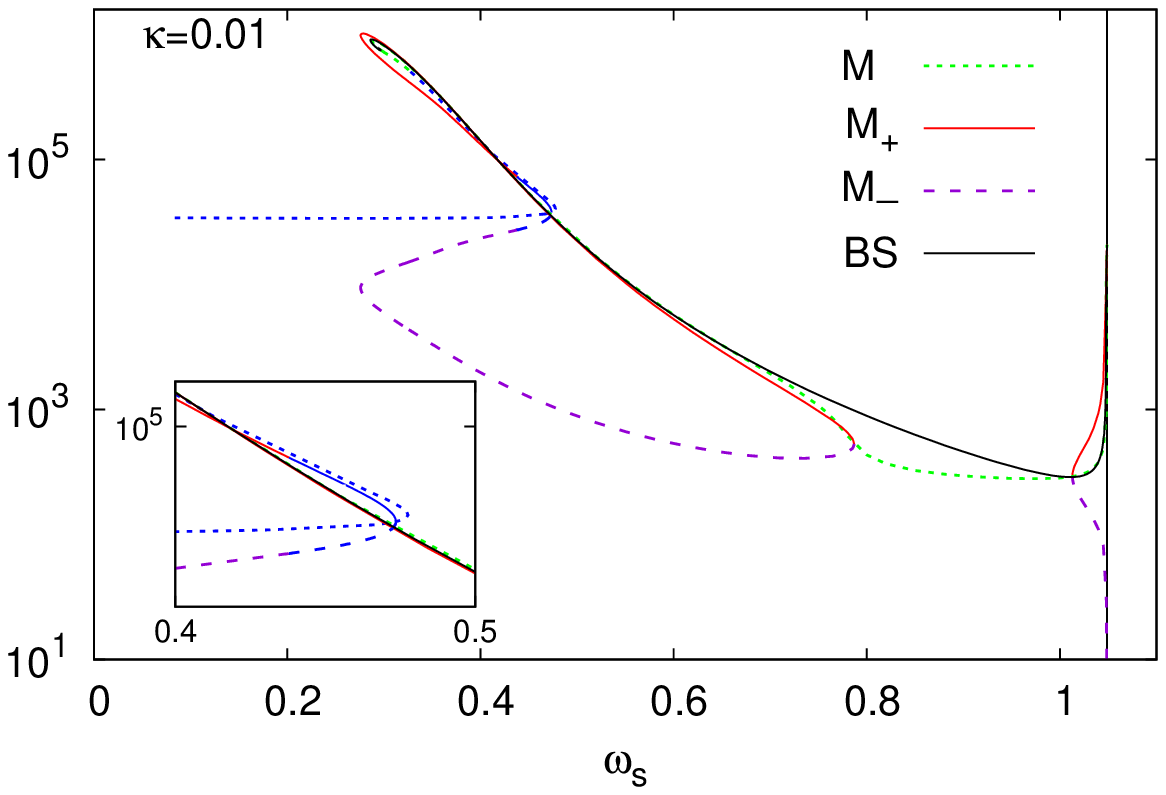}
\label{fig3c}
}
\subfigure[][]{\hspace{-0.5cm}
\includegraphics[height=.25\textheight, angle =0]{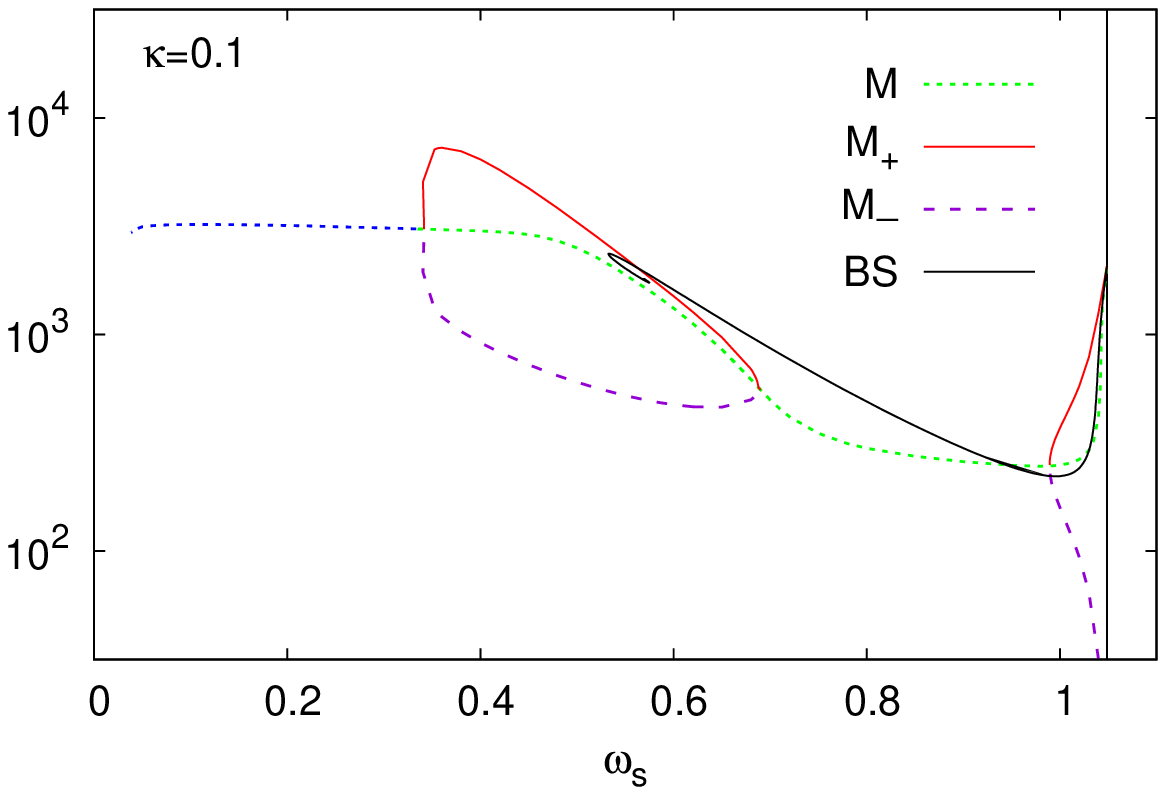}
\label{fig3d}
}
}
\end{center}
\vspace{-0.7cm}
\caption{
Properties of gravitating solutions versus
the boson frequency $\omega_s$ ($\eta_0=3$).
(a) $\kappa=0.001$: The mass $M$ of the symmetric solutions (dotted green), 
and the masses
$M_+$ (solid red) and $M_-$ (dashed lilac)
of the asymmetric solutions.
For double throat configurations the colour is changed to blue.
(b) Same as (a) for the particle number $Q$.
(c) Same as (a) for $\kappa=0.01$
(d) Same as (a) for $\kappa=0.1$.
Also indicated are the masses of the respective boson stars (solid black).
The mass and particle number in the probe limit 
are shown in (a) and (b) for $\omega_s>0.6$ (dotted black).
The thin vertical lines indicate $\omega_{\rm max}$.
\label{fig3}
}
\end{figure}

Let us begin by considering the spontaneous symmetry breaking
in the presence of gravity.
In Fig.~\ref{fig3} we illustrate the dependence of the families of
gravitating solutions on the coupling constant $\kappa$,
keeping the throat parameter $\eta_0$ fixed.
We start with a rather small value of the coupling constant $\kappa$.
In particular, we exhibit in Fig.~\ref{fig3a}
the masses versus the boson frequency for
$\kappa=0.001$ and in Fig.~\ref{fig3b} the respective
particle numbers. We then increase the coupling to 
$\kappa=0.01$ and $\kappa=0.1$, where the respective masses are
shown in Fig.~\ref{fig3c} and Fig.~\ref{fig3d}.

First of all we observe that the symmetry breaking persists
when gravity is coupled.
Thus besides the family of symmetric solutions
also the families of asymmetric solutions are retained.
The bifurcation frequencies
$\omega_{\rm crl}$ and $\omega_{\rm cru}$,
where the pairs of asymmetric solutions bifurcate from 
the symmetric solutions,
vary only slightly with increasing $\kappa$.
As in the probe limit, in between these two bifurcation frequencies
only symmetric solutions exist.

Clearly, in a large range of boson frequencies,
the structure of the gravitating solutions 
follows the structure of the solutions in the probe limit.
In fact, as seen in Fig.~\ref{fig3a} and \ref{fig3b},
the solutions possess very similar global charges there.
However, for very large frequencies and for small frequencies
significant deviations occur.
In particular,
a third bifurcation frequency $\omega_{\rm crg} < \omega_{\rm crl}$
arises, where the asymmetric solutions merge again with the
symmetric ones. This bifurcation is highlighted in the inset
in Fig.~\ref{fig3c}.
This new gravity induced bifurcation at boson frequency $\omega_{\rm crg}$
marks the endpoint of the asymmetric branches.

Let us now go into more detail.
For small $\kappa$ a very prominent new feature
is the spiral structure at large values of the mass
and particle number.
This structure is present for both symmetric and
asymmetric solutions, and resembles at first glance
very much the spiral structure of boson stars.
Indeed, the branches of solutions of topologically
non-trivial solutions follow closely the
branch of boson stars up into the spiral.
However, unlike boson stars, the present spirals unwind again
as seen in the inset in Fig.~\ref{fig3a}.

From a physical point of view,
we consider that the most relevant branches of solutions are
those that start from the minimal value of $M$ (${\cal S}$) or from
the bifurcation frequency $\omega_{\rm crl}$ (${\cal A}_\pm$)
and continue to smaller frequencies until
(for the ${\cal S}$ and ${\cal A}_+$ solutions)
the mass and the particle number reach a maximum.
At this maximum a change of the stability 
of the solutions is expected to occur, in analogy to boson stars.
Namely the solutions should acquire an (additional) unstable mode.
By symmetry the ${\cal A}_-$ would of course also
change stability at the same frequency.

As the spirals unwind, the symmetric and asymmetric configurations
change their geometry and evolve a double throat structure,
as indicated by the change of colour (blue) in the figures.
This will be discussed in more detail below.
Here we note that the families of solutions then
bend backwards and form descending branches
with respect to their global charges, which follow 
closely those of the physically more relevant ascending branches.
When the descent is slowed down, the solutions approach
the third bifurcation frequency $\omega_{\rm crg}$,
where the asymmetric solutions merge again with the
symmetric ones.
Only this symmetric branch then continues 
to smaller frequencies $\omega_s$.

We remark, that for frequencies very close to the maximum frequency 
also significant deviations from the probe limit arise, 
not discerned in Fig.~\ref{fig3a}. 
However, the inset of Fig.~\ref{fig3b}, where the particle number
is shown, reveals that some interesting further branch structure
is present close to $\omega_{\rm max}$ for these small values
of the coupling $\kappa$.
In general, the dependence of the particle number on the frequency
follows very closely the dependence of the mass.
We therefore exhibit the particle number only for a single coupling
constant $\kappa$.
The only noticeable exception to this behaviour occurs for
$\omega_s < \omega_{\rm crg}$ on the last part of the branch of
symmetric solutions (independent of $\kappa$).
Here the particle number increases, 
whereas the mass remains almost constant.

To address the $\kappa$-dependence of the solutions, 
we exhibit in Fig.~\ref{fig3} also the masses for
$\kappa=0.01$ (c) and $\kappa=0.1$ (d).
For $\kappa=0.01$ the spiral structure is still
present at a rudimentary level. Thus the general
structure of the solutions is analogous to the one
of the lower couplings.
When $\kappa$ is further increasing, however, the spiral
behaviour disappears, and the branch structure
of the families of solutions radically simplifies. 
This also holds for frequencies close to the
maximal frequency.
The mass and the particle number of a given family
are then simple functions of the boson frequency $\omega_s$,
since the backbendings present for lower $\kappa$ disappear,
as illustrated for $\kappa=0.1$.

We have focussed our discussion here on the fundamental solutions.
But we would like to remark, that the families of 
first excitations follow the pattern
of the fundamental solutions.


\begin{figure}[t!]
\begin{center}
\vspace{0.5cm}
\mbox{\hspace{-0.5cm}
\subfigure[][]{\hspace{-1.0cm}
\includegraphics[height=.25\textheight, angle =0]{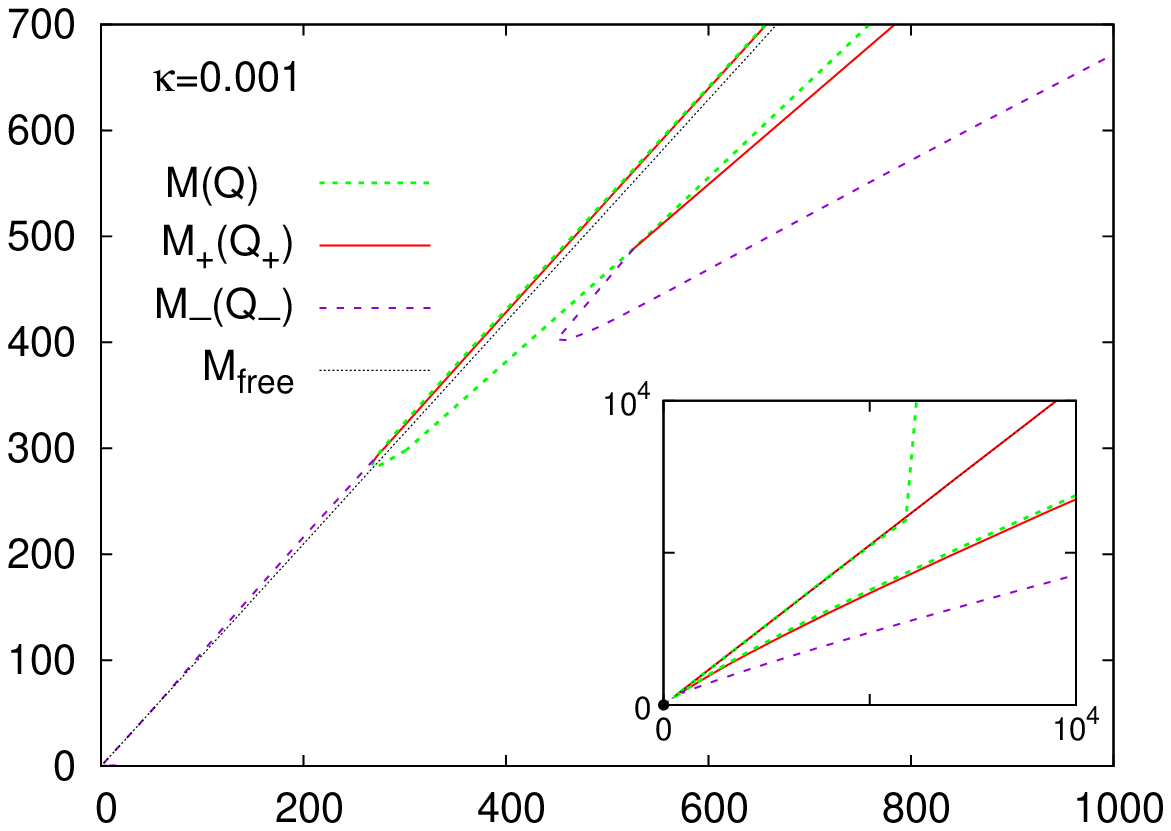}
\label{fig4a}
}
\subfigure[][]{\hspace{-0.5cm}
\includegraphics[height=.25\textheight, angle =0]{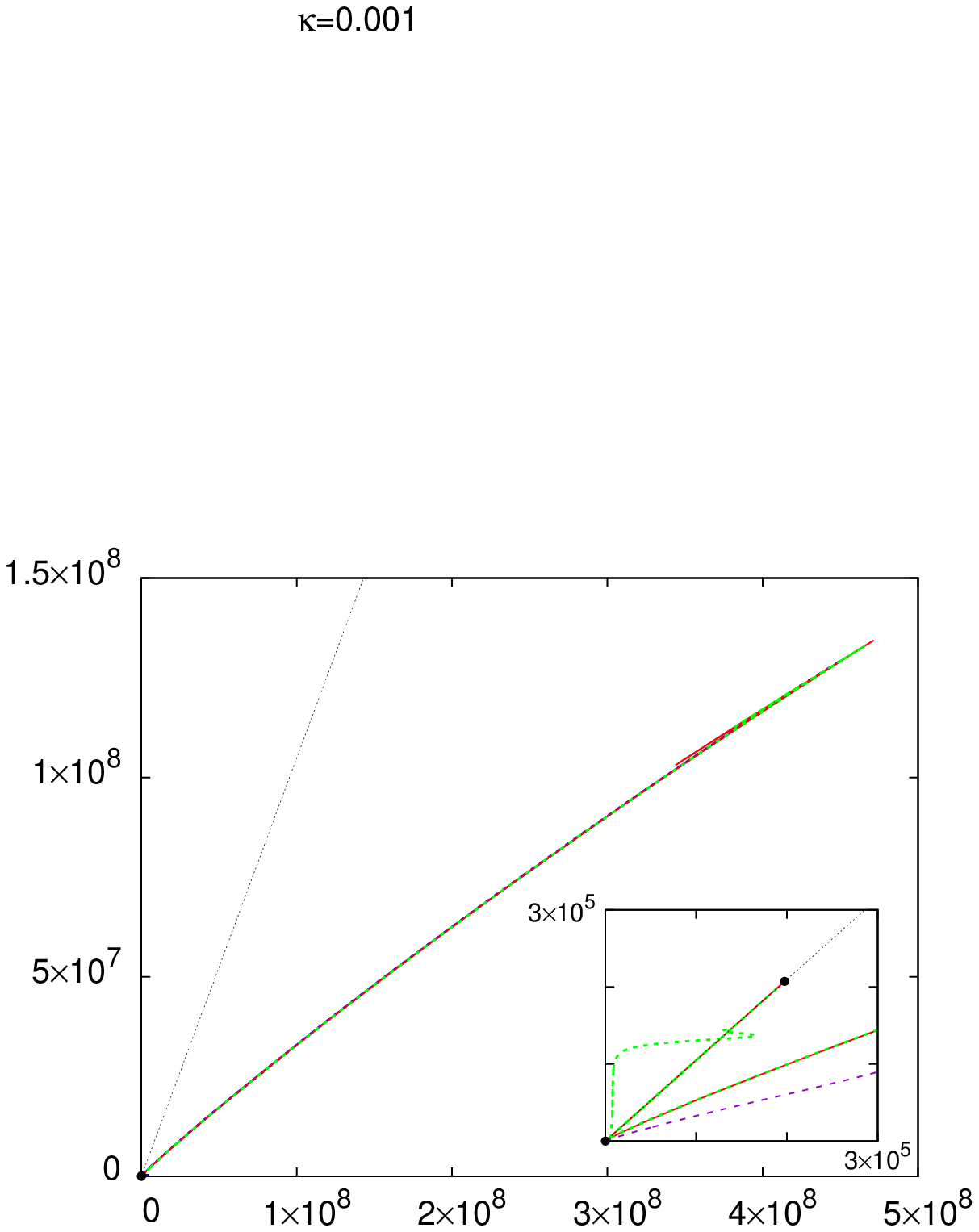}
\label{fig4b}
}
}
\mbox{\hspace{-0.5cm}
\subfigure[][]{\hspace{-1.0cm}
\includegraphics[height=.25\textheight, angle =0]{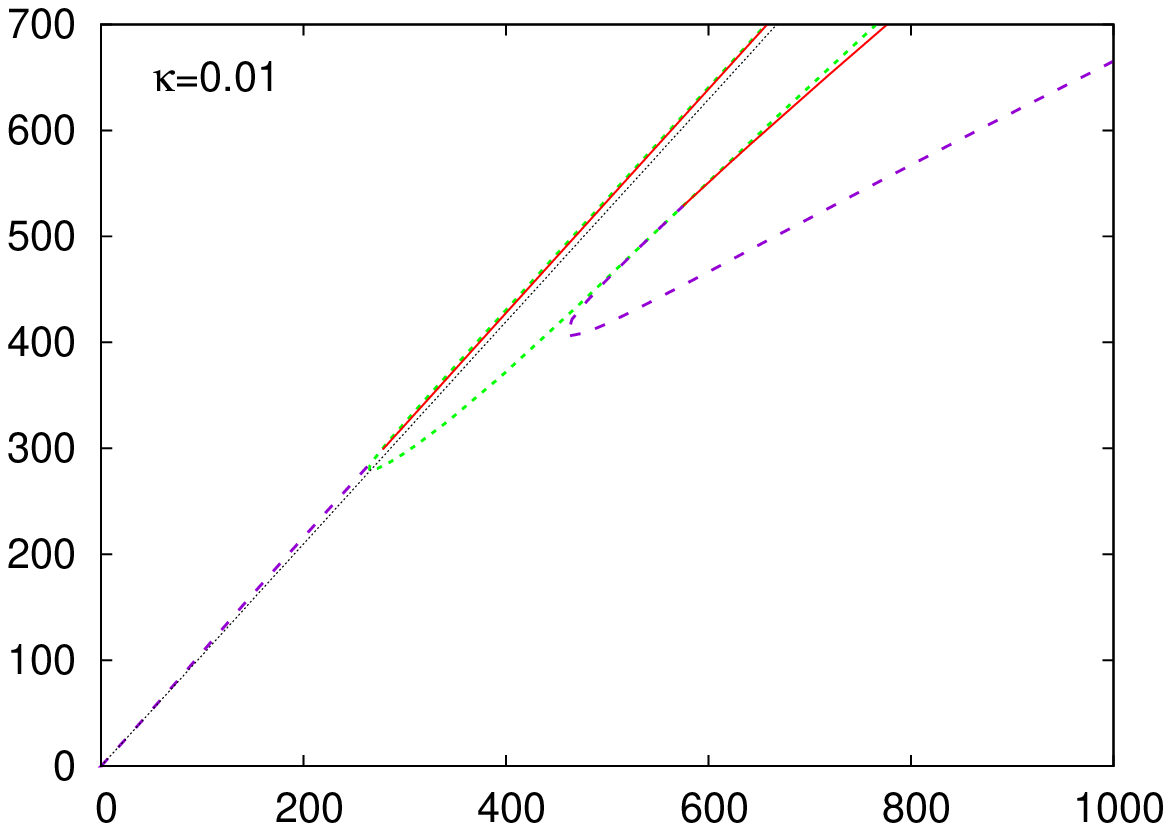}
\label{fig4c}
}
\subfigure[][]{\hspace{-0.5cm}
\includegraphics[height=.25\textheight, angle =0]{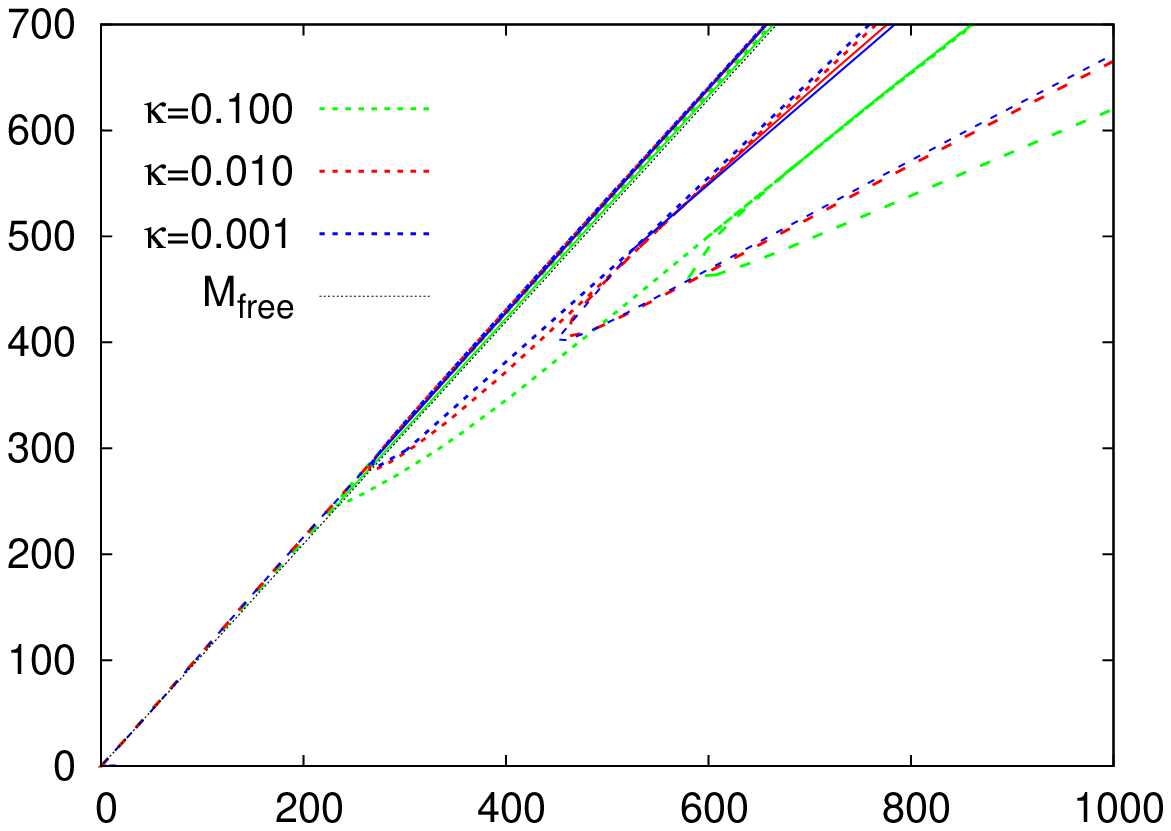}
\label{fig4d}
}
}
\end{center}
\vspace{-0.7cm}
\caption{Gravitating solutions for throat parameter $\eta_0=3$:
(a) The mass $M$ (dotted) of the symmetric solutions
and the mass $M_+$ (solid) and $M_-$ (dashed)
of the asymmetric solutions versus the 
the particle number $Q$, $Q_+$, and $Q_-$, respectively,
for coupling $\kappa=0.001$
in the vicinity of the bifurcations $\omega_{\rm crl}$
and $\omega_{\rm cru}$.
(b) Same as (a) for the full range of masses.
(c)  Same as (a) for $\kappa=0.01$.
(d) Same as (a) for $\kappa=0.1$, $\kappa=0.01$ and $\kappa=0.001$.
Note the different colour coding here.
In all figures the mass of $Q$ free bosons is shown for comparison
(thin black lines).
The black dot in the inset in (b) denotes the
endpoint at $\omega_{\rm max}$.
\label{fig4}
}
\end{figure}

Let us now consider the $M(Q)$ dependence in order
to demonstrate that 
the asymmetric solutions are energetically favoured
also in the presence of gravity.
To this end we exhibit in Fig.~\ref{fig4} the masses
$M$, $M_+$ and $M_-$ 
versus the respective particle numbers 
$Q$, $Q_+$ and $Q_-$
for several values of the coupling constant $\kappa$.
Also shown is the mass of $Q$ free bosons for comparison.
Fig.~\ref{fig4a} presents the masses 
versus the particle numbers 
for the small coupling $\kappa = 0.001$,
zooming into the small mass region,
where the bifurcations at $\omega_{\rm crl}$ and
$\omega_{\rm cru}$ are clearly visible.

As expected, 
for these small masses and particle numbers 
there is very little difference to the probe limit,
except for an overall scaling parameter
(compare Fig.~\ref{fig2d}).
Thus the asymmetric solutions are clearly energetically favoured,
when they bifurcate from the symmetric ones.

But the inset in Fig.~\ref{fig4a} shows already a part
of the additional structure of the
symmetric solutions close to $\omega_{\rm max}$.
The full structure of the solutions is shown in
Fig.~\ref{fig4b}. Interestingly, the various branches of $M(Q)$
for the symmetric and asymmetric solutions are all very close
to each other on such a large scale. The large mass solutions
are all strongly bound as a comparison with the free case shows.
The spiral structure gives rise to a number of cusps at 
large masses. However, since the masses and particle numbers
are all very close, we cannot discern these cusps in the
figure. (Here only a schematic plot would clearly exhibit the cusp
structure.)

The inset in Fig.~\ref{fig4a} shows already a part
of the additional structure of the
symmetric solutions close to $\omega_{\rm max}$.
The black dot in the inset in Fig.~\ref{fig4b}
denotes the endpoint reached by the configurations
at $\omega_{\rm max}$.
This endpoint is universal, i.e., independent of the coupling
$\kappa$.
It also represents a fourth bifurcation point, since 
the asymmetric and symmetric solutions merge again
at $\omega_{\rm max}$.

In Fig.~\ref{fig4c} we demonstrate the
$M(Q)$ dependence for $\kappa=0.01$ in the small mass region.
The figure shows that
the asymmetric solutions remain energetically preferred,
when the coupling is increased.
We observe that the figures \ref{fig4}(a)-(c) look very much the same.
This is demonstrated in \ref{fig4d} for all values of $\kappa$ considered.
Indeed, there is only a slight dependence on the coupling $\kappa$
apart from the overall scaling factor.
Clearly, the asymmetric solutions remain preferred.

\subsubsection{Wormhole Geometries}

\begin{figure}[t!]
\begin{center}
\vspace{0.5cm}
\mbox{\hspace{-0.5cm}
\subfigure[][]{\hspace{-1.0cm}
\includegraphics[height=.25\textheight, angle =0]{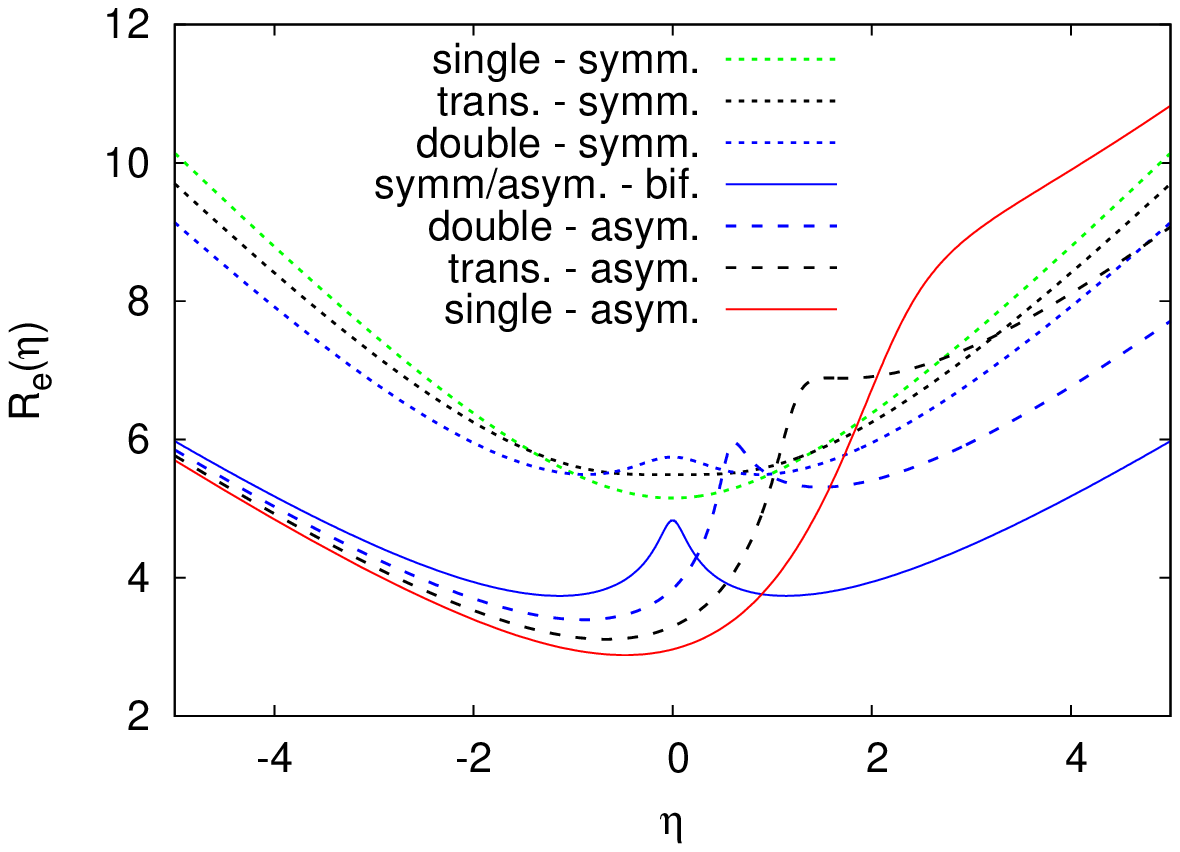}
\label{fig5a}
}
\subfigure[][]{\hspace{-0.5cm}
\includegraphics[height=.25\textheight, angle =0]{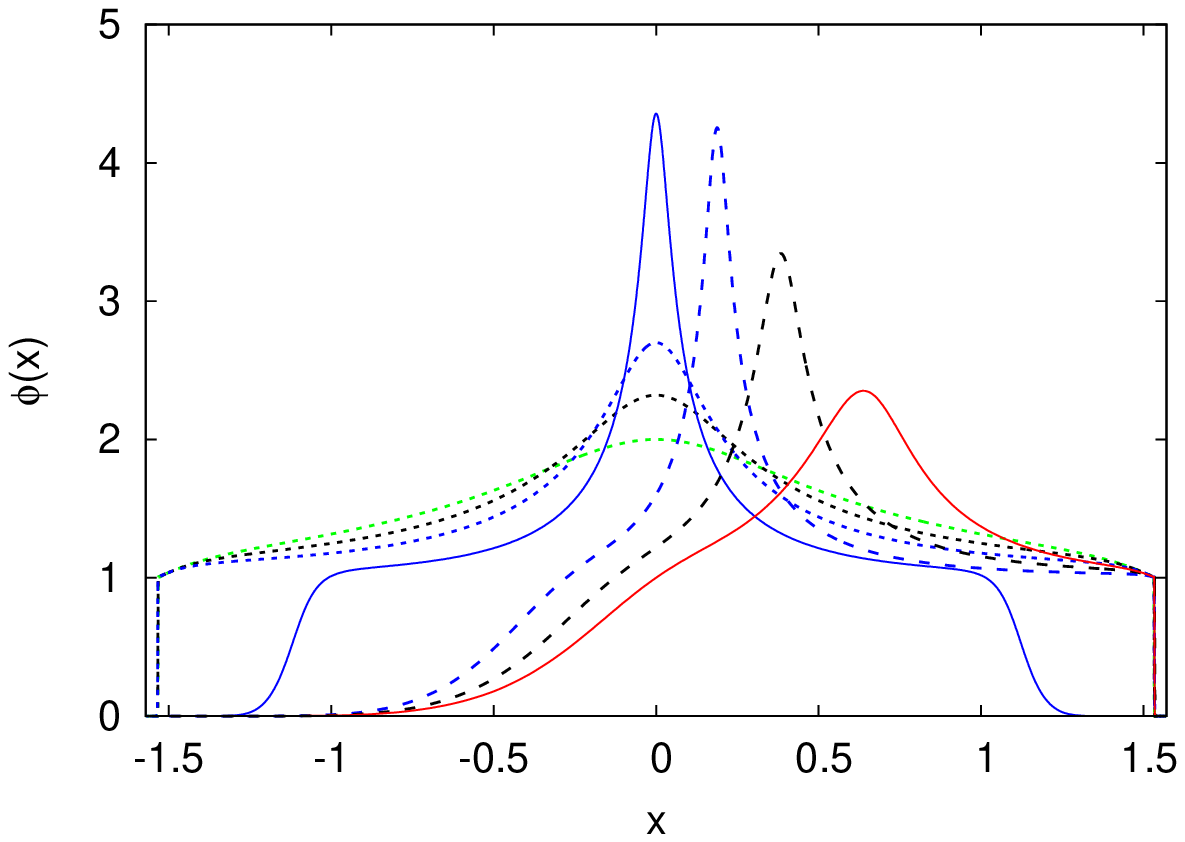}

\label{fig5b}
}
}
\end{center}
\vspace{-0.7cm}
\caption{
Throat structure of non-rotating gravitating solutions ($\eta_0=3$).
(a) The circumferential radius function $R(\eta)$ ($R_+(\eta)$)
versus the compactified coordinate $x={\rm atan}\, \eta$
for a sequence of symmetric 
and asymmetric solutions with various values
of the frequency $\omega_s$ and coupling constant $\kappa = 0.001$.
(b) Same as (a) for the boson function $\phi(\eta)$ ($\phi_+(\eta)$).
\label{fig5}
}
\end{figure}

\begin{figure}[t!]
\begin{center}
\vspace{0.5cm}
\mbox{\hspace{-0.5cm}
\subfigure[][]{\hspace{-1.0cm}
\includegraphics[height=.25\textheight, angle =0]{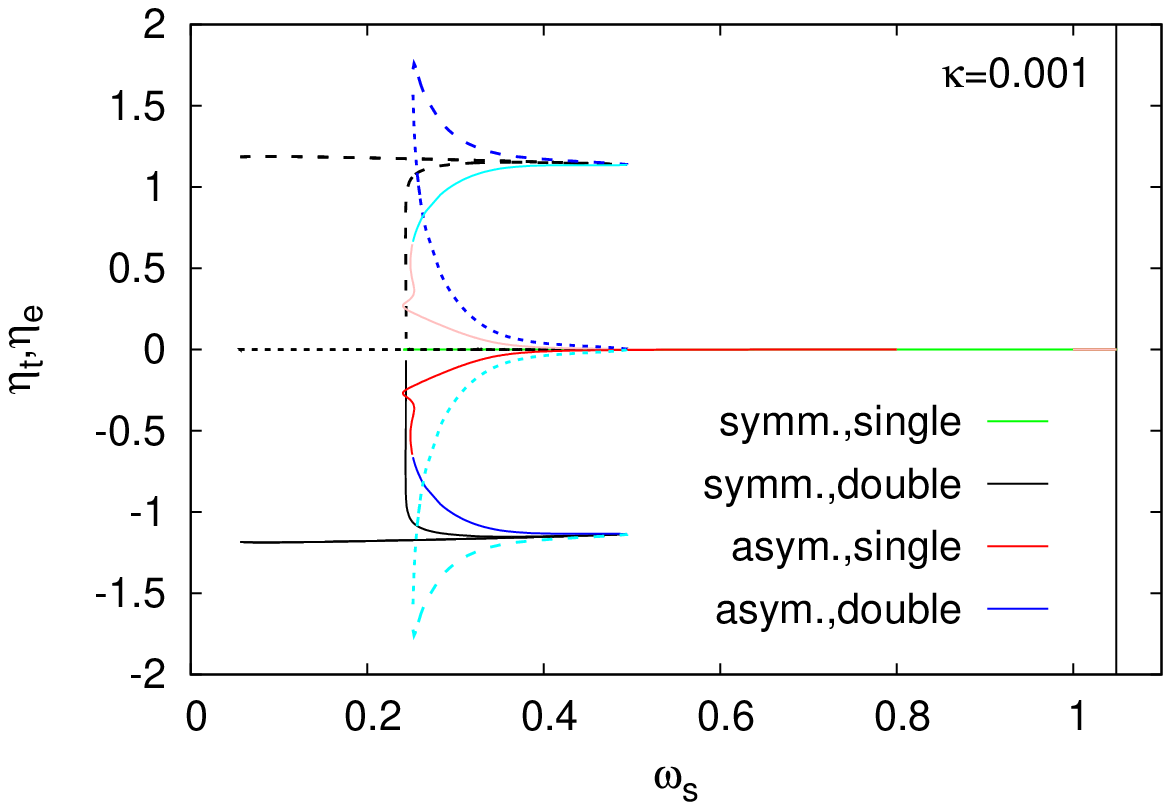}
\label{fig6a}
}
\subfigure[][]{\hspace{-0.5cm}
\includegraphics[height=.25\textheight, angle =0]{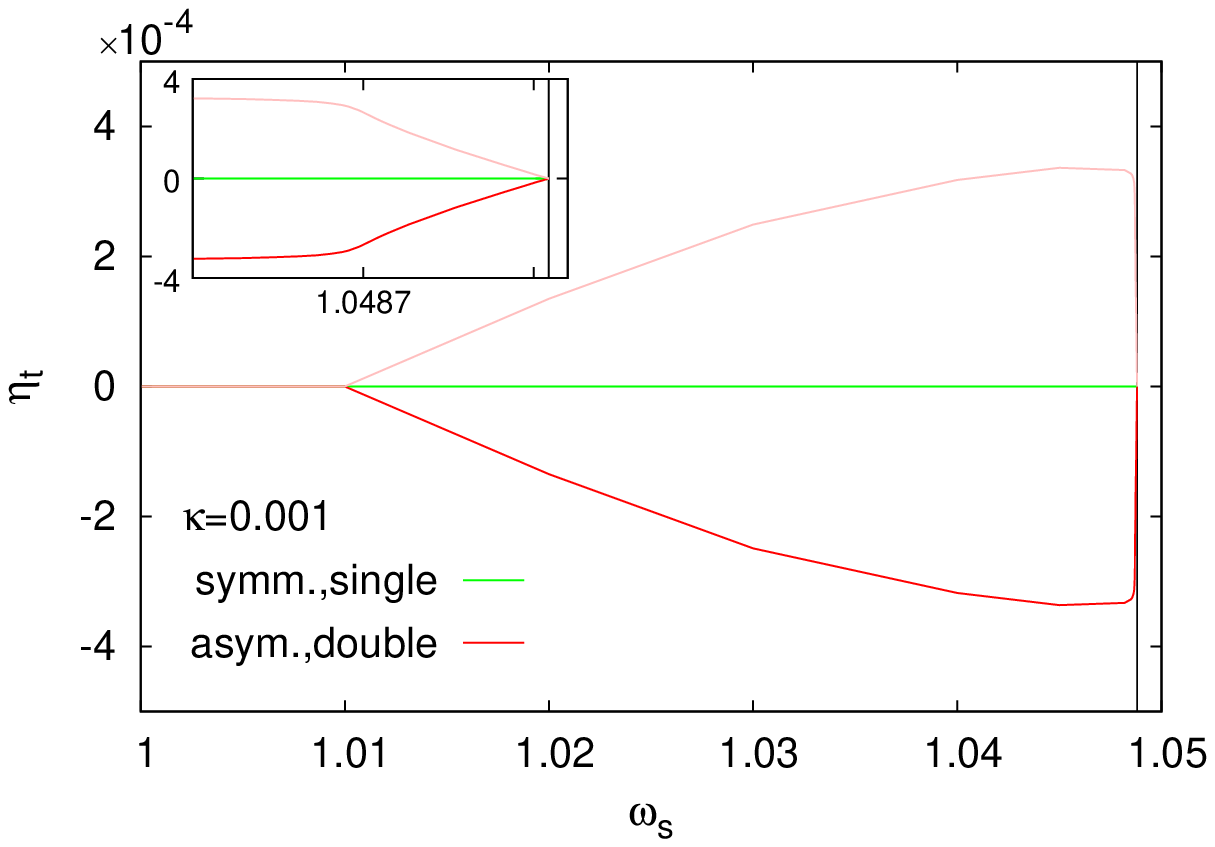}
\label{fig6b}
}
}
\mbox{\hspace{-0.5cm}
\subfigure[][]{\hspace{-1.0cm}
\includegraphics[height=.25\textheight, angle =0]{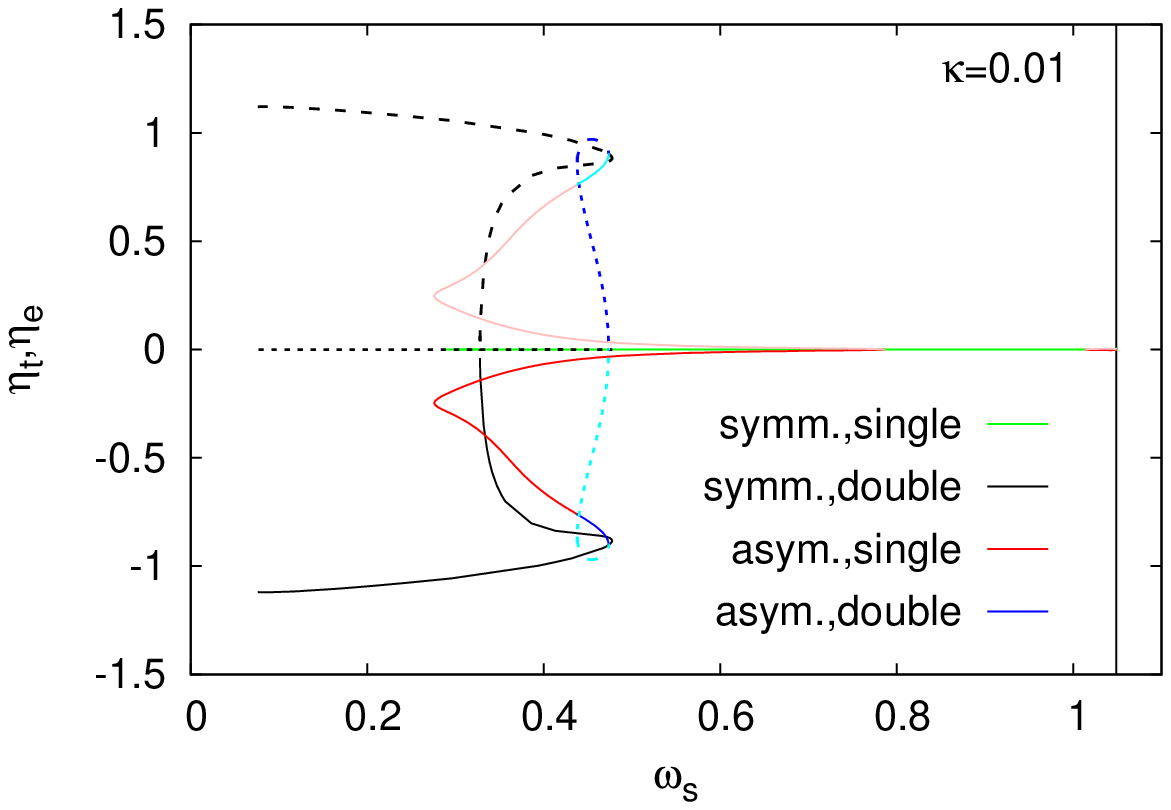}
\label{fig6c}
}
\subfigure[][]{\hspace{-0.5cm}
\includegraphics[height=.25\textheight, angle =0]{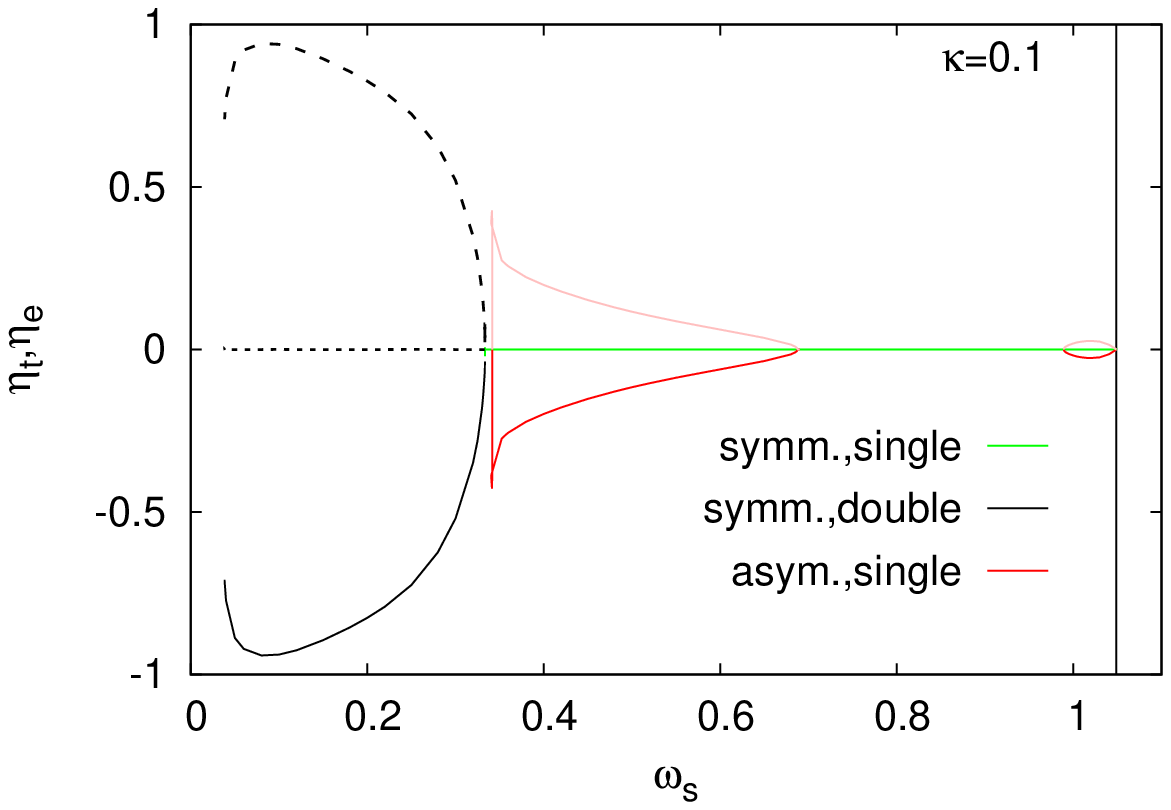}
\label{fig6d}
}
}
\end{center}
\vspace{-0.7cm}
\caption{Gravitating solutions for throat parameter $\eta_0=3$:
(a) Location of the single throat respectively
the double throat and the equator versus the frequency $\omega_s$
for coupling $\kappa = 0.001$.
(b) Zoom of (a) for frequencies close to $\omega_{\rm max}$.
(c) Same as (a) for $\kappa=0.01$.
(d) Same as (a) for $\kappa=0.1$.
\label{fig6}
}
\end{figure}

Next we address the geometry of the solutions.
In particular, we are interested in the transition from single throat
configurations to configurations featuring a double throat 
with an equator in between the two throats.
To illustrate this transition, we exhibit
in Fig.~\ref{fig5a} the circumferential radius function $R(\eta)$
($R_+(\eta)$)
for a sequence of symmetric (asymmetric) solutions,
where the coupling constant has the value $\kappa=0.001$.
Here a minimum of $R(\eta)$ ($R_+(\eta)$)
corresponds to a throat, and
a maximum to an equator. The transition from a single
throat to a double throat configuration occurs at an
inflection point. 

For the symmetric solutions we display in Fig.~\ref{fig5a} a
single throat solution for the frequency $\omega_s=0.2456$,
the transition from a single to a double throat
configuration happening at $\omega_{\rm tr}=0.2439$,
and a double throat solution with $\omega_s=0.2435$.
The  asymmetric double throat solutions bifurcate at $\omega_{\rm crg}=0.4942$
from the symmetric ones.
For the asymmetric solutions the respective solutions have frequencies
$\omega_s=0.2724$ (double throat), $\omega_{\rm tr}=0.2518$ and $\omega_s=0.2494$
 (single throat).
As seen in the figure, in the asymmetric case the equator and
the second throat emerge asymmetrically.
Thus 
the single throat does not degenerate at the transition
frequency, where the equator and the second throat arise.
Instead, an inflection point arises in the other part
of the manifold, which splits into a maximum and a minimum as the
frequency $\omega_s$ is increased.

In addition we display in the figure
the double throat solution at the third
bifurcation frequency $\omega_{\rm crg}$,
where the symmetric and asymmetric solutions merge again.
To get an idea of the matter distributions associated with
these configurations and, in particular, with the transitions, 
we exhibit in Fig.~\ref{fig5b} the boson function $\phi(\eta)$ 
($\phi_+(\eta)$) for the
same set of solutions.
We observe that the value of $\phi_0$ 
increases from the single to the double throat solutions.
In particular, we note that the inflection point arises 
close to the peak of $\phi_+(\eta)$
in the part of the manifold, where most of the matter resides.

To see the evolution of the throats of the symmetric and
asymmetric configurations, we exhibit the
dependence of their locations in Fig.~\ref{fig6},
beginning in Fig.~\ref{fig6a} with the coupling $\kappa=0.001$.
Clearly,
for symmetric solutions their single throat (solid green)
is localized at $\eta_t=0$,
while beyond their transition frequency $\omega_{\rm tr}$
their equator resides at $\eta_e=0$ (dotted black).
At the transition frequency the double throats (solid and dashed 
black) emerge and stride away.

For the larger frequencies the single throat 
of the asymmetric solutions is located close to $\eta_t=0$.
However, for the solutions ${\cal A}_+$ it is then shifted into
the region ${\cal M}_-$ (and vice versa) (solid dark red),
because of the backreaction of the matter on the geometry.
At the transition frequency this throat continues to
stride away (solid dark blue), while a cusp arises
in the region ${\cal M}_+$, formed by the second throat
(dashed dark blue) and the equator (dotted dark blue).
Note that the light colours (red and blue) in the figure
represent the second asymmetric solution.
At the bifurcation $\omega_{\rm crg}$, 
the asymmetric throats merge with the symmetric throats,
and the asymmetric equator with the symmetric one.

In Fig.~\ref{fig6b} we zoom into the region close to $\omega_{\rm max}$.
Here only single throat solutions exist. The inset demonstrates 
the merging of the solutions at $\omega_{\rm max}$.
We note that in this region the asymmetric throats first stride far away
from $\eta=0$, before they return towards the limiting symmetric solution.

As $\kappa$ is increased, the frequency range of 
the asymmetric double throat solutions decreases, 
as seen in Fig.~\ref{fig6c} for $\kappa = 0.01$, 
where the same colour coding is used. 
For large $\kappa$ only symmetric double throat solutions
are left, as illustrated in Fig.~\ref{fig6d} for $\kappa=0.1$.
These symmetric double throat solutions persist
at small frequencies for any value of the coupling $\kappa$.

\begin{figure}[t!]
\begin{center}
\vspace{-0.5cm}
\mbox{\hspace{-0.5cm}
\subfigure[][]{\hspace{-1.0cm}
\includegraphics[height=.25\textheight, angle =0]{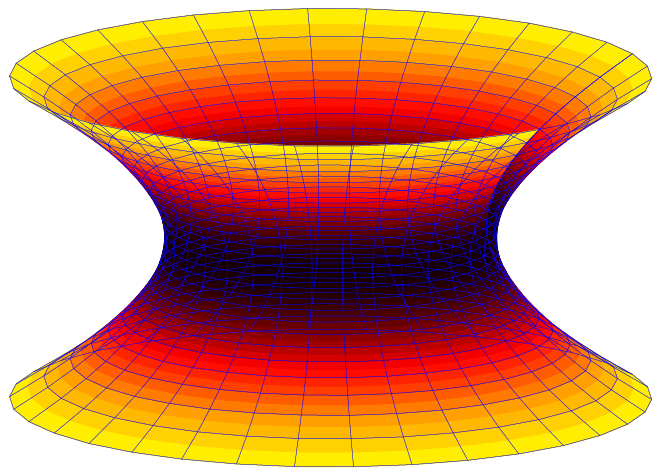}
\label{fig7a}
}
\subfigure[][]{\hspace{-0.5cm}
\includegraphics[height=.25\textheight, angle =0]{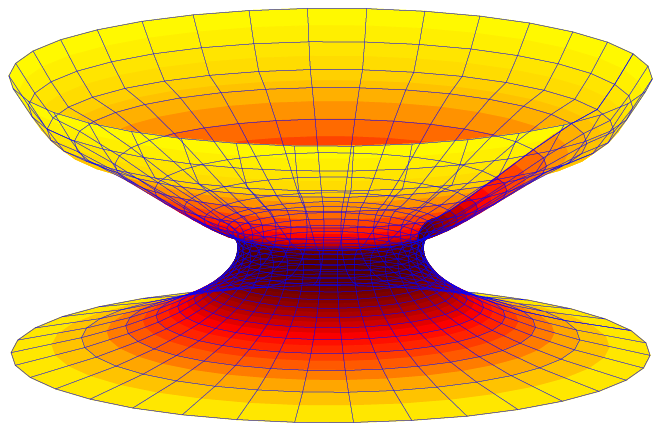}
\label{fig7b}
}
}
\mbox{\hspace{-0.5cm}
\subfigure[][]{\hspace{-1.0cm}
\includegraphics[height=.25\textheight, angle =0]{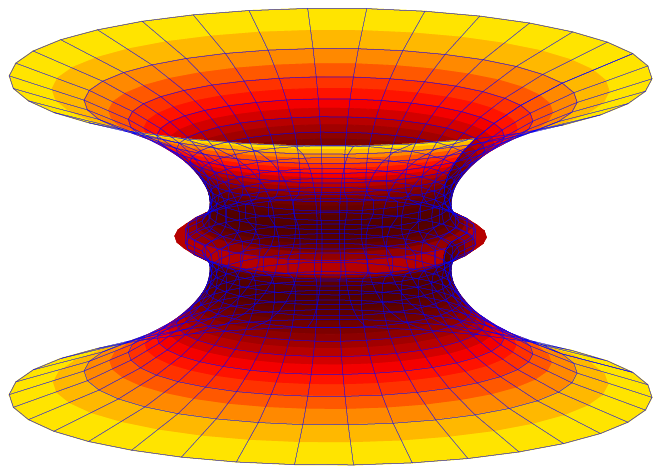}
\label{fig7c}
}
\subfigure[][]{\hspace{-0.5cm}
\includegraphics[height=.25\textheight, angle =0]{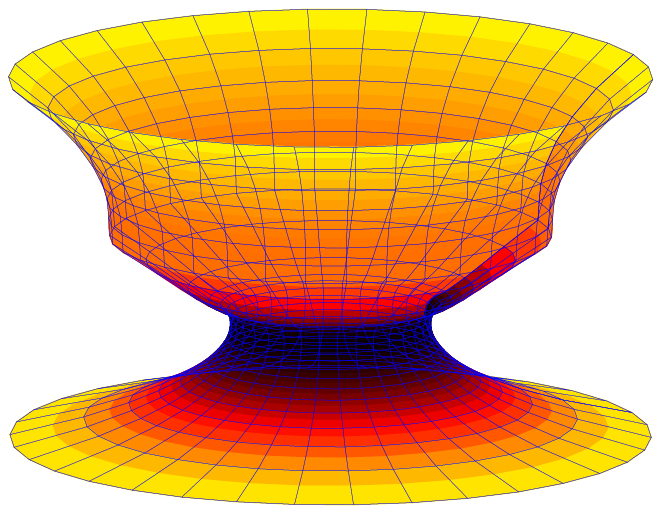}
\label{fig7d}
}
}
\end{center}
\vspace{-0.7cm}
\caption{
Throat geometry:
Three dimensional view of the isometric embedding.
(a) Symmetric solution with a single throat for
$\kappa=0.001$, $\omega=0.3$.
(b) Asymmetric solution with a single throat for
$\kappa=0.001$, $\omega=0.3$.
(c) Symmetric solution with a double throat for
$\kappa=0.001$, $\omega=0.3$.
(d) Asymmetric solution with a double throat for
$\kappa=0.001$, $\omega=0.3$.
\label{fig7}
}
\end{figure}

Let us now visualize the geometry of the wormholes by 
considering embeddings of spatial hypersurfaces.
We display an isometric embedding of one of the equatorial planes
($\theta=0,\ \pi/2$)
for several wormhole spacetimes in Fig.~\ref{fig7}.
For the embedding we employ the parametric representation
\begin{equation}
\rho(\eta) = R(\eta) \ , \ \ \ \ 
z(\eta)= \int_0^\eta\sqrt{1-R'^2} \ d\eta' \ .
\label{embedd}
\end{equation}

Below the transition value $\omega_{\rm tr}$
the solutions possess a single throat, as seen in 
Fig.~\ref{fig7} for a symmetric (a) and 
an asymmetric (b) solution.
For the parameters chosen in the figure,
$\kappa=0.001$ and $\omega_s=0.3$,
the single throat of the asymmetric solution
has been pushed into ${\cal M}_-$.
The respective double throat solutions are shown
in Fig.~\ref{fig7}(c) and (d).
For the symmetric ones the equator resides at the centre (c),
while for the asymmetric ones it is located
off the centre (d).
We did not find solutions with more than two throats.


Finally, we would like to briefly comment on the limiting solution
for small boson frequencies.
As discussed above, for small $\omega_s$ only symmetric solutions
persist.
When $\omega_s$ is decreased here, the boson function becomes
steeper and steeper in the vicinity of the equator,
with its peak strongly increasing.
At the same time the circumferential function $R(\eta)$
peaks more and more strongly at the equator.
This behaviour indicates that a singular limiting solution is reached,
whose Kretschmann scalar diverges in the limit.
This agrees with the previous four-dimensional study
\cite{Dzhunushaliev:2014bya}.

\subsection{Spontaneous Symmetry Breaking in Four Spacetime Dimensions}

\begin{figure}[t!]
\begin{center}
\vspace{0.5cm}
\mbox{\hspace{-0.5cm}
\subfigure[][]{\hspace{-1.0cm}
\includegraphics[height=.25\textheight, angle =0]{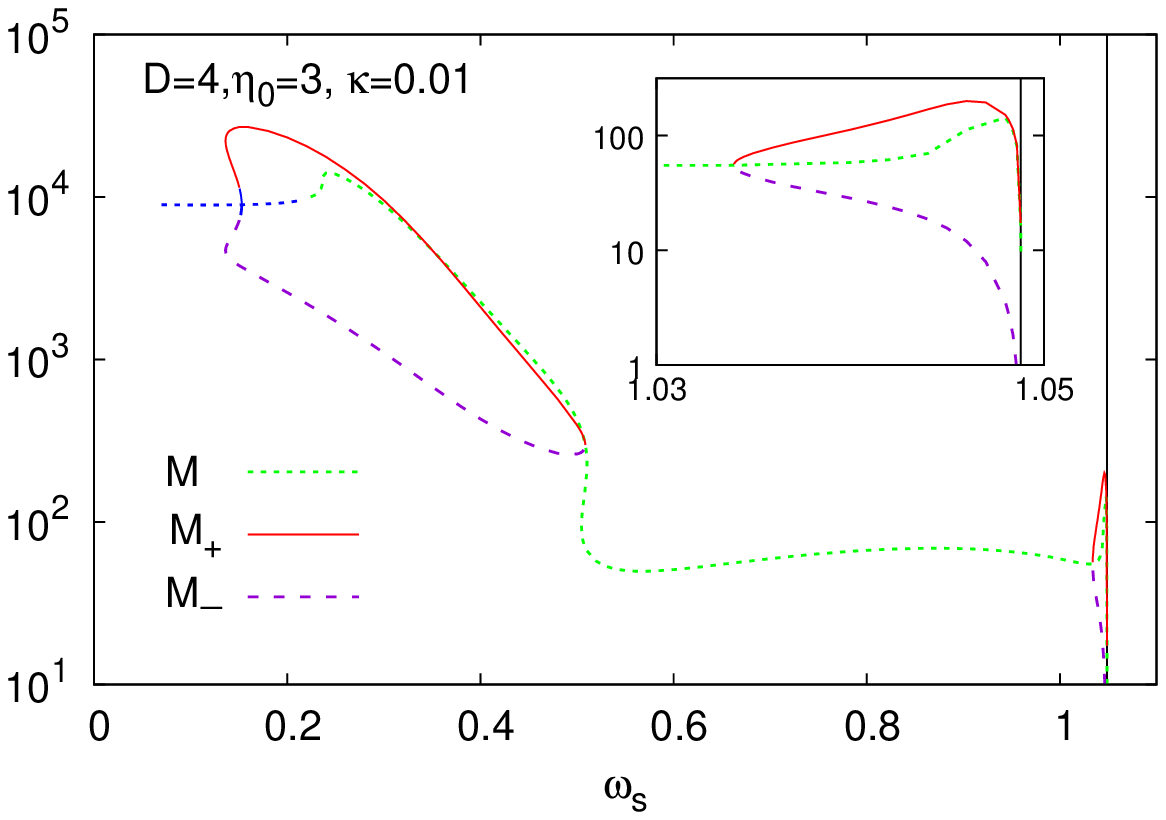}
\label{fig8a}
}
\subfigure[][]{\hspace{-0.5cm}
\includegraphics[height=.25\textheight, angle =0]{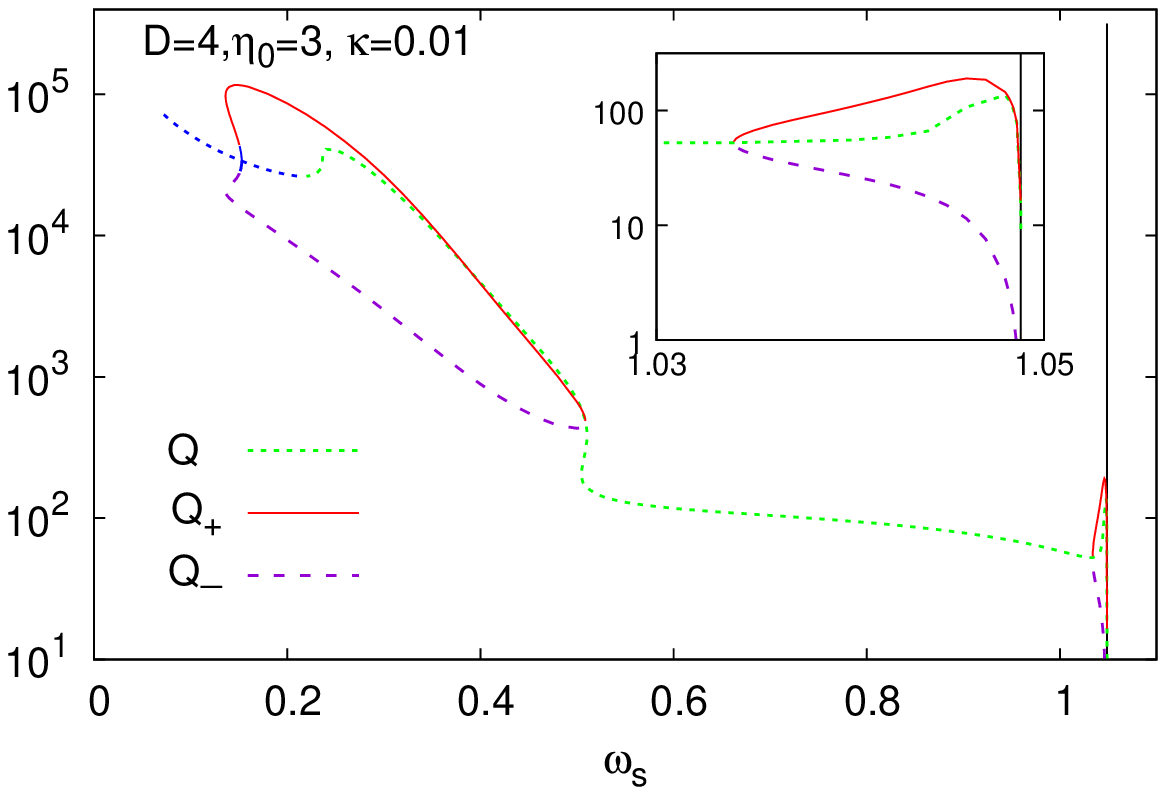}
\label{fig8b}
}
}
\mbox{\hspace{-0.5cm}
\subfigure[][]{\hspace{-1.0cm}
\includegraphics[height=.25\textheight, angle =0]{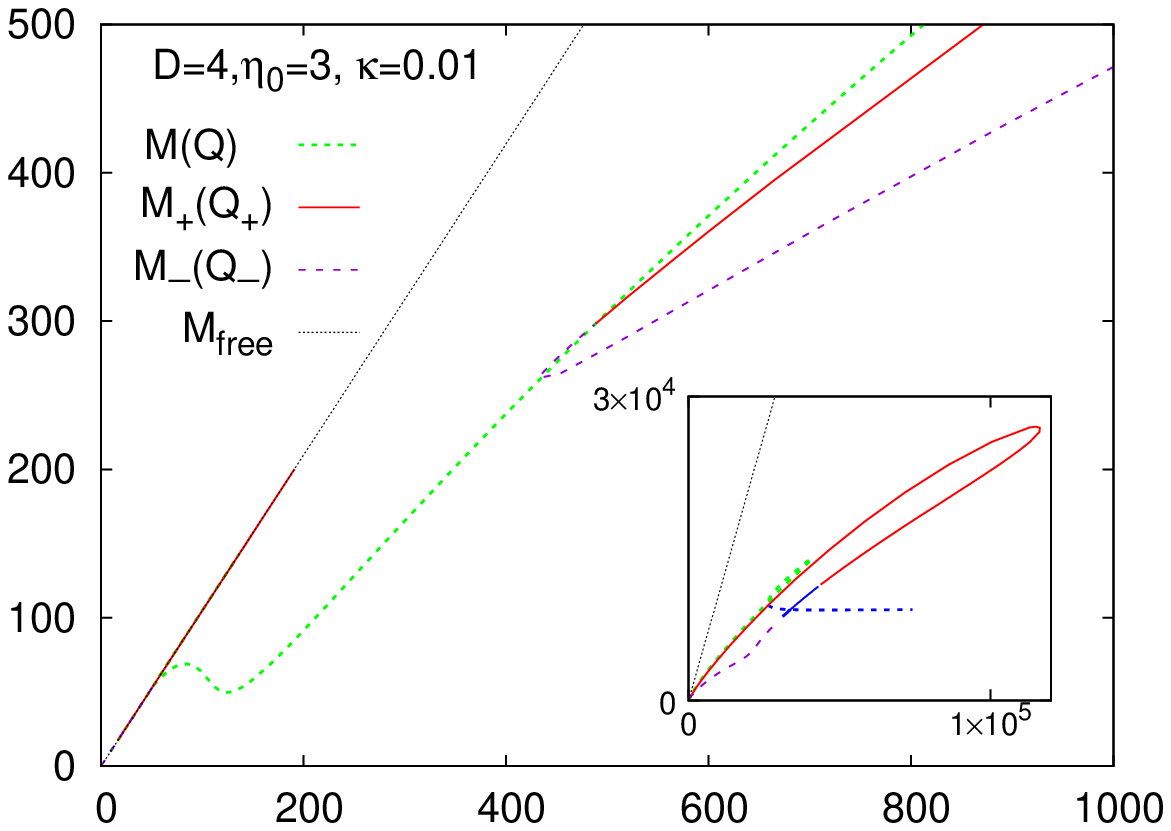}
\label{fig8c}
}
\subfigure[][]{\hspace{-0.5cm}
\includegraphics[height=.25\textheight, angle =0]{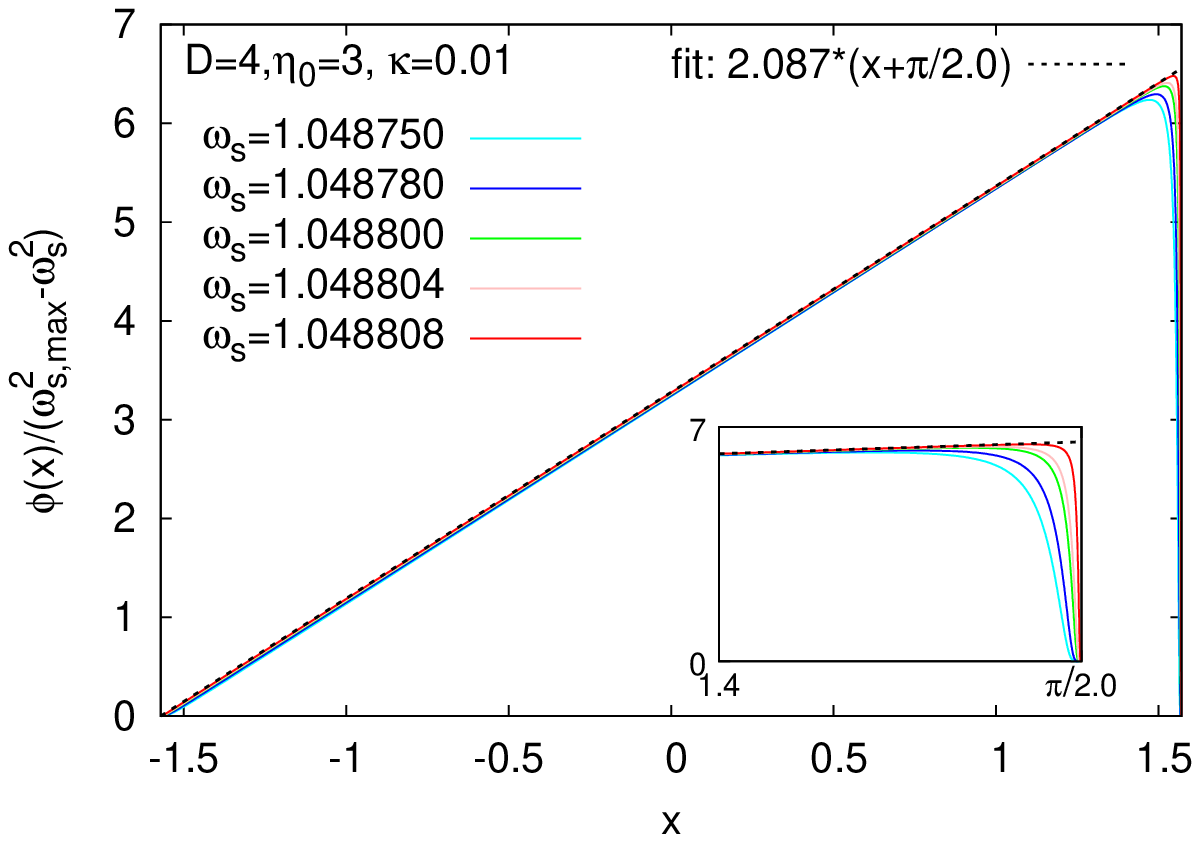}
\label{fig8d}
}
}
\end{center}
\vspace{-0.7cm}
\caption{
Properties of gravitating solutions versus
the boson frequency $\omega$ ($\eta_0=3$) in four dimensions.
(a) $\kappa=0.01$: The mass $M$ of the symmetric solutions (dotted green),
and the masses
$M_+$ (solid red) and $M_-$ (dashed lilac)
of the asymmetric solutions.
For double throat configurations the colour is changed to blue.
(b) Same as (a) for the particle number $Q$.
(c) The mass $M$ (dotted) of the symmetric solutions
and the mass $M_+$ (solid) and $M_-$ (dashed)
of the asymmetric solutions versus the
the particle number $Q$, $Q_+$, and $Q_-$, respectively,
for coupling $\kappa=0.01$
in the vicinity of the bifurcations $\omega_{\rm crl}$
and $\omega_{\rm cru}$.
(d) The limit $\omega_s \to \omega_{\rm max}$ is demonstrated 
for the boson field function $\phi$.
\label{fig8}
}
\end{figure}

After having discussed in detail the properties of the
wormholes within bosonic matter in five spacetime dimensions,
we now briefly demonstrate that the features observed
in five dimensions also hold in four dimensions.
We recall, that
in \cite{Dzhunushaliev:2014bya} only the symmetric solutions
have been studied, which we now complement with the asymmetric ones.

First of all we note, that the analogous bifurcation phenomenon
takes place in four dimensions. 
In the probe limit,
below a critical value $\eta_{0\, \rm cr}$ of the throat parameter,
symmetric and asymmetric solutions exist in the full 
frequency interval $[\omega_{\rm min}, \omega_{\rm max}]$.
Above this critical value a frequency gap
$\omega_{\rm crl} \le \omega_s \le \omega_{\rm cru}$
appears, where only symmetric solutions are found.
At the end points of the gap, i.e., at $\omega_{\rm cru}$
and $\omega_{\rm crl}$,
pairs of asymmetric solutions bifurcate from the symmetric ones,
and persist for $\omega_{\rm min} < \omega_s <\omega_{\rm crl}$
and $\omega_{\rm cru} < \omega_s < \omega_{\rm max}$.

As in five dimensions,
this bifurcation phenomenon is retained in the presence of gravity.
%
Depending on the coupling constant $\kappa$, the branch structure
of the solutions changes considerably, as already demonstrated in 
\cite{Dzhunushaliev:2014bya} for the symmetric solutions
(and for a smaller throat parameter).
The asymmetric solutions follow this general pattern.
Thus with increasing $\kappa$ the branch structure simplifies,
and the spiral part disappears,
analogously to the five-dimensional solutions.
We here demonstrate the branch structure
for the value of $\kappa=0.01$.

Fig.~\ref{fig8a} exhibits the masses 
$M$ (dotted green),
$M_+$ (solid red) and $M_-$ (dashed lilac),
where the bifurcations are clearly visible.
The asymmetric solutions emerge from the symmetric ones at
the bifurcations $\omega_{\rm crl}$ and $\omega_{\rm cru}$,
and merge again at
$\omega_{\rm crg}$ (as well as at $\omega_{\rm max}$).
Also, the transition from single to double throat solutions
is observed (blue)
analogously to the five-dimensional case.

The respective particle numbers are shown in Fig.~\ref{fig8b}.
Again, they follow the behaviour of the masses rather closely,
except for very small frequencies $\omega_s$, where only
the symmetric solutions are retained.
In Fig.~\ref{fig8c}, the masses $M$, $M_+$ and $M_-$
are shown versus the respective particle numbers
$Q$, $Q_+$ and $Q_-$ in the bifurcation region
(i.e., for the larger values of the frequency $\omega_s$),
together with the mass of $Q$ free bosons.
The inset in Fig.~\ref{fig8c} exhibits 
the solutions in their full range of existence.

As expected, the asymmetric solutions are energetically 
favoured over the symmetric ones. Thus the spontaneous
symmetry breaking leads to more strongly bound systems.
We conjecture, that the analogous phenomenon is present
also in more than five dimensions.
Let us remark, that
in four dimensions we are in principle entitled to also address
astrophysical aspects of these solutions.
Then one may view them as 
boson stars with a nontrivial topology.
However, this aspect will be addressed elsewhere.

As a final point, we remark that in four dimensions
the limiting solution for $\omega_s \to \omega_{\rm max}$
becomes rather trivial in the sense, that the
boson field tends to zero and the solution has vanishing
mass. This is different for five dimensions,
where the limiting solution has a non-trivial boson field
and a finite mass.
We exhibit in Fig.~\ref{fig8d} a sequence of solutions
close to $\omega_{\rm max}$, which demonstrates
the way the boson field approaches its limit.

\section{Conclusions and Outlook}

In this paper we have considered Ellis wormholes in the presence of
a complex bosonic matter field and encountered the phenomenon
of spontaneous symmetry breaking of the solutions.
Starting with the probe limit, we have seen that besides
the symmetric configurations there are also asymmetric configurations
present.
The latter always come in pairs, and are related to each other by
a reflection with respect to the radial coordinate $\eta=0$.

For a small throat size, the symmetric and asymmetric solutions
are present in the full frequency interval
[$\omega_{\rm min}$,$\omega_{\rm max}$], whereas for a larger
throat size this is only true for the symmetric solutions.
There the asymmetric solutions branch off the symmetric ones
at critical values of the frequency, 
$\omega_{\rm crl}$ and $\omega_{\rm cru}$.
At the critical value of the throat size $\eta_{0\,  \rm cr}$
these critical frequency values coincide.

Both symmetric and asymmetric solutions satisfy the same
set of boundary conditions. In that sense
the asymmetric solutions appear spontaneously,
without any external trigger. The reason for their appearance
is that the asymmetric solutions are energetically
favourable, as we have shown.

To this end, we have analyzed the masses and particle numbers
of the solutions. For the symmetric solutions one finds the
same mass $M$ in both asymptotic regions, and likewise
the same particle number.
For the asymmetric solutions this is different. Here
the solution with most of its mass located in the region ${\cal M}_+$
possesses the mass $M_+$ in this region 
and the mass $M_-$ in ${\cal M}_-$. For the second asymmetric
solution of the pair these masses are interchanged
because the two are related by reflection.
The same holds for the particle number.

When the mass of the three solutions
is then considered versus their respective particle number,
it becomes clear, that asymmetric solutions are
more strongly bound than the symmetric solutions.
Thus the spontaneous symmetry breaking leads to
energetically favoured configurations.

All of this remains valid when gravity is coupled.
However, gravity introduces further new features.
First of all,
the backreaction of the boson field on the
metric removes the lower frequency bound $\omega_{\rm min}$.
For small values of the coupling constant $\kappa$
a spiralling behaviour arises, that is known from compact stars.
However, unlike for compact stars, here the spirals
unwind again. We attribute this effect to the presence
of the negative energy density in the form of a
phantom field, since such unwinding has, for instance, also been observed
for bosonic configurations in Einstein-Gau\ss -Bonnet theory 
\cite{Hartmann:2013tca}.

Second, the presence of gravity generates a further bifurcation
phenomenon at $\omega_{\rm crg}$. Here the pair of asymmetric 
solutions merges again with the symmetric ones.
Only the symmetric solutions then persist to small boson frequencies.
Third, we observe a transition in the geometry of the
solutions. For small $\kappa$ this transition arises
in the vicinity of the spiral. Here the single throat solutions
develop an equator and a second throat. For the symmetric
solutions, the equator is then localized at the radial coordinate
$\eta=0$, surrounded symmetrically by both throats.
For the asymmetric solutions, in contrast, the equator and
the second throat originate far from the first throat 
in the other part of the manifold.

Here we have performed most of the calculations in 
five spacetime dimensions. However, the formalism
is general for $D$ dimensions. Pure Ellis wormholes
in $D$ dimensions have been obtained in \cite{Torii:2013xba}.
It should be straightforward to include bosonic
matter and obtain the analogues of the
configurations studied here also in $D>5$ dimensions.
In particular, we expect that the phenomenon of
spontaneous symmetry breaking will be present
independent of the dimension.

We have already shown that this phenomenon is also
present in four dimensions. In this case,
such configurations of bosonic matter
surrounding wormholes might also be of 
potential astrophysical interest \cite{Dzhunushaliev:2014bya}.
In particular,
one can obtain solutions which differ widely in mass and size 
by varying the potential for the complex scalar field.
For instance, there may also be solutions which
mimic compact astrophysical objects like neutron stars
or black holes.
These solutions could be studied in the context 
of gravitational lensing
\cite{Abe:2010ap,Toki:2011zu,Takahashi:2013jqa},
with respect to their light curves \cite{Dzhunushaliev:2014mza}, 
their geodesics
\cite{Eilers:2013lla}, etc.

Let us conclude with some remarks on the stability of these
solutions.
The Ellis wormholes are known to be unstable
\cite{Shinkai:2002gv,Gonzalez:2008wd,Gonzalez:2008xk,Torii:2013xba}.
The unstable radial mode of the wormholes was shown to
persist in the presence of bosonic matter
for symmetric solutions in four dimensions
\cite{Dzhunushaliev:2014bya},
where the instability was weakened by the presence of matter.

Now, that we have seen, that spontaneous symmetry breaking occurs,
we conjecture that the symmetric solutions will acquire
another unstable mode from the matter side. In contrast,
the asymmetric solutions will only possess the single
unstable wormhole mode (on their fundamental branch).
While we defer a full analyis of the stability of the symmetric and
asymmetric solutions to a later time,
we note already, that the analysis of the solutions in the
probe limit precisely conforms to this expectation.
Here, at the bifurcation frequencies
$\omega_{\rm crl}$ and $\omega_{\rm cru}$
the symmetric solutions indeed exhibit a zero mode,
which then turns into a second unstable mode
in the parameter space where the asymmetric solutions exist.
For the asymmetric solutions
stability could possibly be achieved
by removing the phantom field
and modifying gravity instead 
\cite{Hochberg:1990is,Fukutaka:1989zb,Ghoroku:1992tz,Furey:2004rq,Bronnikov:2009az,Kanti:2011jz,Kanti:2011yv,Lobo:2009ip,Harko:2013yb}.

\section*{Acknowledgement}

We gratefully acknowledge support by the German Research Foundation
within the framework of the DFG Research Training Group 1620
{\it Models of gravity}
as well as support by the Volkswagen Stiftung, and support from FP7, Marie Curie
Actions, People, International Research Staff Exchange
Scheme (IRSES-606096).
BK
gratefully acknowledges support from Fundamental Research in Natural Sciences
by the Ministry of Education and Science of Kazakhstan.

\section{Appendix}

We briefly explain, how we obtain the mass 
and the particle number of the asymmetric solutions.
For these special care is needed, since there is some ambiguity
as to where to put the lower limit of the respective volume integrals.
We therefore extract these global charges from the
asymptotic behaviour of the solutions.
For the symmetric solutions there is no such ambiguity.

\subsection{Mass in the probe limit}

In the probe limit the backreaction of the matter on the 
spacetime is not taken into account, since the matter equation
is solved in the background of the Ellis wormhole.
Then the mass cannot be extracted from the asymptotic form of the metric.
To obtain the mass anyway asymptotically, we resort to the 
following construction.

We consider the Einstein equation
\begin{equation}
R^0_0\sqrt{-g} = \kappa \left(T^0_0 +\frac{1}{2-D}T_\mu^\mu\right)\sqrt{-g}
\end{equation}
and treat the metric function $a$ as order ${\cal O}(\kappa)$. Consequently,
only the background metric enters the right hand side.
Evaluation yields
\begin{equation}
-\frac{D-3}{2}\left(
\left[\left(ph\right)^{\frac{D-2}{2}}\frac{1}{\sqrt{p}}\right]a'\right)'
=
\kappa \left(T^0_0 +\frac{1}{2-D}T_\mu^\mu\right)
\left(ph\right)^{\frac{D-2}{2}}\sqrt{p} \ .
\end{equation}
Defining $\rho_{\rm m} = \left[\left(ph\right)^{\frac{D-2}{2}}\frac{1}{\sqrt{p}}\right]a'$
we find 
\begin{equation}
\rho_{\rm m}(\eta) = -\frac{2}{D-3} \kappa
\int_{-\infty}^\eta {  \left(T^0_0 +\frac{1}{2-D}T_\mu^\mu\right)
                      \left(ph\right)^{\frac{D-2}{2}}\sqrt{p} } d\eta' 
		      + \rho_{\rm m 0} \ ,
\end{equation}
where $\rho_{\rm m \ 0}$ is an integration constant. In the next step we integrate
$a' =\left[\left(ph\right)^{\frac{2-D}{2}}\sqrt{p}\right] \rho_{\rm m}$,
\begin{eqnarray}
a(\eta)  & = & -\frac{2\kappa}{D-3}\int_{-\infty}^\eta {
\left[\left(ph\right)^{\frac{2-D}{2}}\sqrt{p}\right]
\left\{
\int_{-\infty}^{\eta'} {  \left(T^0_0 +\frac{1}{2-D}T_\mu^\mu\right)
                      \left(ph\right)^{\frac{D-2}{2}}\sqrt{p} } d\eta''
\right\} } d\eta'
\nonumber\\
& &
+ \rho_{\rm m 0}\int_{-\infty}^\eta {\left[\left(ph\right)^{\frac{2-D}{2}}\sqrt{p}\right]} d\eta'
+a_0 \ ,
\end{eqnarray}
where $a_0$ is again an integration constant.
Next we find the integration constants $\rho_{\rm m 0}$ and $a_0$ from 
the boundary conditions $a(\pm \infty)=0$,
\begin{equation}
\rho_{\rm m 0}=
\frac{2\kappa}{D-3}
\frac{
\int_{-\infty}^\infty {
\left[\left(ph\right)^{\frac{2-D}{2}}\sqrt{p}\right]
\left\{
\int_{-\infty}^{\eta} {  \left(T^0_0 +\frac{1}{2-D}T_\mu^\mu\right)
                      \left(ph\right)^{\frac{D-2}{2}}\sqrt{p} } d\eta'
\right\} } d\eta 
} 
{\int_{-\infty}^\infty {\left[\left(ph\right)^{\frac{2-D}{2}}\sqrt{p}\right]} d\eta}
\ , \ \ a_0=0 \ .
\end{equation}
The masses $M_\pm$ are related to the asymptotic behaviour of the function $a$,
\begin{equation}
M_\pm = \pm \frac{D-3}{2\kappa}\left[\eta^{D-2} a'\right]_{\pm \infty}\Omega_{D-2} 
=  \pm \frac{D-3}{2\kappa}\rho_{\rm m}(\pm \infty)\Omega_{D-2} \ .
\end{equation}
Explicitly,
\begin{eqnarray}
M_+ & = & \left\{-\frac{D-2}{D-3}\int_{-\infty}^\infty{\left(T^0_0 +\frac{1}{2-D}T_\mu^\mu\right)
                                                    \left(ph\right)^{\frac{D-2}{2}}\sqrt{p} } d\eta
	   +\frac{D-2}{2\kappa} \rho_{\rm m 0}\right\}\Omega_{D-2} \ , 
\nonumber\\
M_- & = & 	-\frac{D-2}{2\kappa}\rho_{\rm m 0}\Omega_{D-2} \ .
\end{eqnarray} 	
We note that $M_+$ can be written as
\begin{equation}
M_+ =	\frac{D-2}{D-3}	\int_\Sigma {
\left(T_{\mu\nu} +\frac{1}{2-D} g_{\mu\nu}T_\lambda^\lambda\right)n^\mu \xi^\nu }dV
+\frac{D-2}{2\kappa} \rho_{\rm m 0}\Omega_{D-2} \ ,
\end{equation}
where $\Sigma$ denotes a spacelike hypersurface including both asymptotic regions
of ${\cal M}_+$ and  ${\cal M}_-$.	              
\subsection{Particle number via electric charge}

We consider a fictitious electrostatic potential $\Phi_{\rm el}(\eta)$
sourced by the current $j^\mu$, Eq.~(\ref{current}),
\begin{equation}
\partial_\mu\left(
\sqrt{-g} g^{\mu \nu} g^{t t}\partial_\nu \Phi_{\rm el}
\right)  
 = 
 -j^t \sqrt{-g} \ .
\end{equation}
For the spherically symmetric Ansatz this yields
\begin{equation}
\left(e^{(D-3)a}\left(ph\right)^{\frac{D-2}{2}}\frac{1}{\sqrt{p}}
      \Phi'_{\rm el}\right)' 
= j^t e^{-a}\left(ph\right)^{\frac{D-2}{2}}\sqrt{p}
\ .
\label{Maxw}
\end{equation}
The charges $Q_{\pm}$ can then be obtained from 
\begin{equation}
Q_{\pm} = 
\mp \left[\
\eta^{D-2} \Phi'_{\rm el}
\right]_{\pm\infty} \Omega_{D-2} \ .
\end{equation}
Introducing the auxiliary quantity 
$\rho_q= e^{(D-3)a}\left(ph\right)^{\frac{D-2}{2}}\frac{1}{\sqrt{p}}
      \Phi'_{\rm el}$
we obtain from Eq.~(\ref{Maxw})
\begin{equation}
\rho_q(\eta) = 
\int_{-\infty}^\eta {j^t e^{-a}\left(ph\right)^{\frac{D-2}{2}}\sqrt{p}} d\eta'
+\rho_{q 0}
\end{equation}
with integration constant $\rho_{q 0}$. On the other hand, expressing 
$\Phi'_{\rm el}$ in terms of $\rho_q$ and integrating yields
\begin{equation}
\Phi_{\rm el}(\eta)=
\int_{-\infty}^\eta{
e^{-(D-3)a}\left(ph\right)^{-\frac{D-2}{2}}\sqrt{p}
\left[
\int_{-\infty}^{\eta'} {j^t e^{-a}\left(ph\right)^{\frac{D-2}{2}}\sqrt{p}} d\eta''
\right]}
d\eta'
+\rho_{q 0}\int_{-\infty}^\eta{e^{-(D-3)a}\left(ph\right)^{-\frac{D-2}{2}}\sqrt{p}}
d\eta' + \Phi_{\rm el  0} \ ,
\end{equation}
where $\Phi_{\rm el  0}$ is another integration constant. 
In order to detemine  the integration constants $\rho_{q 0}$ and $\Phi_{\rm el  0}$ we
impose the boundary conditions $\Phi_{\rm el}(\pm\infty)=0$.
This yields $\Phi_{\rm el  0}=0$ and 
\begin{equation}
\rho_{q 0}=
-\frac{
\int_{-\infty}^\infty{
e^{-(D-3)a}\left(ph\right)^{-\frac{D-2}{2}}\sqrt{p}
\left[
\int_{-\infty}^\eta {j^t e^{-a}\left(ph\right)^{\frac{D-2}{2}}\sqrt{p}} d\eta'
\right]}
d\eta}{
\int_{-\infty}^\infty{e^{-(D-3)a}\left(ph\right)^{-\frac{D-2}{2}}\sqrt{p}}
d\eta} \ .
\end{equation}
We observe that $Q_\pm = \mp\rho_q(\pm \infty)\Omega_{D-2}$. Consequently,
\begin{equation}
Q_+ = 
-\left(\int_{-\infty}^\infty {j^t e^{-a}\left(ph\right)^{\frac{D-2}{2}}\sqrt{p}} d\eta
+\rho_{q 0}\right)\Omega_{D-2} \ , \ \ 
Q_- = \rho_{q 0}\Omega_{D-2} \ .
\end{equation}
\end{document}